\RequirePackage{xcolor,ifpdf}
\documentclass[hyper,letterpaper]{JHEP3}
\pdfoutput=1
\usepackage{tikz}
\usepackage{graphicx}

\usepackage{amsmath,amssymb,amsfonts,bm,empheq}
\usepackage{cite}
\usepackage{subcaption}
\usepackage{nicefrac}
\usepackage[enableskew,vcentermath]{youngtab}
\usepackage{epstopdf}
\usepackage{url}
\usepackage{float}
\usepackage{slashed,framed,xcolor,tikz}
\def\printname#1{
        \if\draft y
                \smash{\makebox[0pt]{\hspace{-0.5in}
                        \raisebox{8pt}{\tt\tiny #1}}}
        \fi
}

\newlength{\standardunitlength}
\catcode`\@=11
\long\def\@makecaption#1#2{%
     \vskip 10pt

\setbox\@tempboxa\hbox{%\ifvoid\tinybox\else\box\tinybox\fi
       \small\sf{\bfcaptionfont #1. }\ignorespaces #2}%
     \ifdim \wd\@tempboxa >\captionwidth {%
         \rightskip=\@captionmargin\leftskip=\@captionmargin
         \unhbox\@tempboxa\par}%
       \else
         \hbox to\hsize{\hfil\box\@tempboxa\hfil}%
     \fi}
\font\bfcaptionfont=cmssbx10 scaled \magstephalf
\newdimen\@captionmargin\@captionmargin=2\parindent
\newdimen\captionwidth\captionwidth=\hsize
\catcode`\@=12

\newdimen\tableauside\tableauside=1.0ex
\newdimen\tableaurule\tableaurule=0.4pt
\newdimen\tableaustep
\def\phantomhrule#1{\hbox{\vbox to0pt{\hrule height\tableaurule width#1\vss}}}
\def\phantomvrule#1{\vbox{\hbox to0pt{\vrule width\tableaurule height#1\hss}}}
\def\sqr{\vbox{%
\phantomhrule\tableaustep
\hbox{\phantomvrule\tableaustep\kern\tableaustep\phantomvrule\tableaustep}%
\hbox{\vbox{\phantomhrule\tableauside}\kern-\tableaurule}}}
\def\squares#1{\hbox{\count0=#1\noindent\loop\sqr
\advance\count0 by-1 \ifnum\count0>0\repeat}}
\def\tableau#1{\vcenter{\offinterlineskip
\tableaustep=\tableauside\advance\tableaustep by-\tableaurule
\kern\normallineskip\hbox
    {\kern\normallineskip\vbox
      {\gettableau#1 0 }%
     \kern\normallineskip\kern\tableaurule}%
  \kern\normallineskip\kern\tableaurule}}
\def\gettableau#1 {\ifnum#1=0\let\next=\null\else
  \squares{#1}\let\next=\gettableau\fi\next}

\newcommand{\bea}{\begin{eqnarray}}
\newcommand{\eea}{\end{eqnarray}}
\newcommand{\be}{\begin{equation}}
\newcommand{\ee}{\end{equation}}

\renewcommand{\bar}{\overline}
\renewcommand{\hat}{\widehat}

\newcommand\catalannumber[3]{
  \fill[white]  (#1) rectangle +(#2,#2);
  \fill[fill=gray!25]
  (#1)
  \foreach \dir in {#3}{
    \ifnum\dir=0
    -- ++(0,1)
    \else
    -- ++(1,0)
    \fi
  } |- (#1);
  \draw[help lines] (#1) grid +(#2,#2);
  \draw[dashed] (#1) -- +(#2,#2);
  \coordinate (prev) at (#1);
  \foreach \dir in {#3}{
    \ifnum\dir=0
    \coordinate (dep) at (0,1);
    \else
    \coordinate (dep) at (1,0);
    \fi
    \draw[line width=2pt,-stealth] (prev) -- ++(dep) coordinate (prev);
  };
}

\usetikzlibrary{calc}

\usepackage{mathtools}
\def\multiset#1#2{\ensuremath{\left(\kern-.1em\left(\genfrac{}{}{0pt}{}{#1}{#2}\right)\kern-.1em\right)}}

\catcode`,\active

\catcode`\,12

\catcode`,\active

\catcode`\,12

\title{Nahm sums, quiver A-polynomials and topological recursion}

\author{H\'elder Larragu\'ivel$^{1}$, Dmitry Noshchenko$^{1}$, Mi{\l}osz Panfil$^{1}$ and Piotr Su{\l}kowski$^{1,2}$
\\
$^1$ Faculty of Physics, University of Warsaw, ul. Pasteura 5, 02-093 Warsaw, Poland \\
$^2$ Walter Burke Institute for Theoretical Physics, California Institute of Technology, Pasadena, CA 91125, USA 
}

\abstract{We consider a large class of $q$-series that have the structure of Nahm sums, or equivalently motivic generating series for quivers. First, we initiate a systematic analysis and classification of classical and quantum A-polynomials associated to such $q$-series. These quantum quiver A-polynomials encode recursion relations satisfied by the above series, while classical A-polynomials encode asymptotic expansion of those series. Second, we postulate that those series, as well as their quantum quiver A-polynomials, can be reconstructed by means of the topological recursion. There is a large class of interesting quiver A-polynomials of genus zero, and for a number of them we confirm the above conjecture by explicit calculations. In view of recently found dualities, for an appropriate choice of quivers, these results have a direct interpretation in topological string theory, knot theory, counting of lattice paths, and related topics. In particular it follows, that various quantities characterizing those systems, such as motivic Donaldson-Thomas invariants, various knot invariants, etc., have the structure compatible with the topological recursion and can be reconstructed by its means. 
\\
\\
\\
\\
\\
\\
\\ 
\\
\\
\\
\\
{\tt CALT-2020-014}}

\begin{document}

\tableofcontents

%%%%%%%%%%%%%%%%%%%%%%%%%%%%%%%%%%%%%%%%%%%%%%%%%%%%%%%%%%%%%%%%%%%%%

\newpage

\section{Introduction}    \label{sec-intro}

There is an interesting class of $q$-series that arise in various context in mathematics and physics, whose structure is encoded in a matrix $C$ with integer entries $C_{i,j}$, and which in addition depend on $m$ generating parameters $x_1,\ldots, x_m$
\begin{equation}
P_C(x_1, \dots, x_m) = \sum_{d_1, \dots, d_m\geq 0}\frac{(-q^{1/2})^{\sum_{i,j=1}^m C_{i,j} d_i d_j}}{(q; q)_{d_1}\cdots (q;q)_{d_m}} x_1^{d_1} \cdots x_m^{d_m}.    \label{PC}
\end{equation}
The sums in the above expression are made over nonnegative integers $d_1,\ldots,d_m$. 

First, such series were found in certain statistical models and identified as characters of coset conformal field theories \cite{Kedem:1993ve,Kedem:1992jv}. 

Second, the series (\ref{PC}) can be interpreted as Nahm sums \cite{Nahm,Zagier2007}. For special choices of $C$ and generating parameters $x_i$ such Nahm sums characterize integrable field theories, however it is also of interest to consider them for more general $C$ and $x_i$. Among others, Nahm sums have interesting modularity properties and are closely related to knot invariants \cite{Garoufalidis2015,garoufalidis2018asymptotics}. 

Third, the matrix $C$ can be interpreted as defining a symmetric quiver, with $C_{i,j}=C_{j,i}$ encoding the number of arrows from vertex $i$ to vertex $j$. In this case the above series is identified as the motivic generating series for the quiver defined by the matrix $C$. A certain product decomposition of this generating series into quantum dilogarithms encodes motivic Donaldson-Thomas invariants associated to this quiver, which characterize a moduli space of representations of this quiver  \cite{Kontsevich:2010px,COM:8276935,Rei12,MR,FR}. 

The series (\ref{PC}) plays also a prominent role in the knots-quivers correspondence formulated in \cite{Kucharski:2017poe,Kucharski:2017ogk}, and analyzed and generalized further in
\cite{Kucharski:2016rlb,Luo:2016oza,Stosic:2017wno,Panfil:2018sis,Ekholm:2018eee,Panfil:2018faz,Ekholm:2019lmb,Stosic:2020xwn}. According to this correspondence, the generating series of knot polynomials, as well as certain open topological string partition functions, can be written in the form (\ref{PC}), for an appropriate choice of $C$. Among others, this enables to express certain open BPS invariants, referred to as the LMOV invariants, in terms of motivic Donaldson-Thomas invariants, thereby providing a long sought after proof of integrality of these former invariants \cite{OoguriV,Labastida:2000zp,Labastida:2000yw,Labastida:2001ts,Ramadevi_Sarkar,Mironov:2017hde}.  
The knots-quivers correspondence relates to each other also various other objects from the realms of knot theory, quiver representation theory, and topological string theory, and thus it provides one of the motivations for this work. Furthermore, as an important byproduct of the knots-quivers correspondence, it turns out that various series of the form (\ref{PC}) capture counts of certain classes of lattice paths \cite{Panfil:2018sis}.

The series (\ref{PC}) has an interesting asymptotic expansion, which can be determined via the saddle point (WKB) approximation, and which is encoded in the following Nahm equations
\begin{equation}  
(-1)^{C_{i,i}}x_i \prod_{j=1}^m z_j^{C_{i,j}} + z_i -1 = 0, \qquad i=1,\ldots, m.  \label{clNahm-intro}
\end{equation} 
It turns out that an interesting information is also encoded in the so called mixed resultant (with respect to $z_i$) of the above equations and an additional equation $y=\prod_{i=1}^m z_i$. The resultant, denoted $A(x_1,\ldots,x_m,y)$, is an irreducible polynomial that vanishes only if Nahm equations and $y=\prod_i z_i$ have a common root. Furthermore, introducing non-commutative operators $\hat{z}_l$ that satisfy the Weyl algebra $\hat{z}_l \hat{x}_k = q^{\delta_{kl}} \hat{x}_k \hat{z}_l$, it turns out that a quantum version Nahm equation holds, which is a statement that the following operators annihilate the generating function (\ref{PC})
\begin{equation}
 \Big( (-q^{1/2})^{C_{k,k}} \hat{x}_k \prod_j^m \hat{z}_j^{C_{kj}} + \hat{z}_k - 1\Big) P_C(x_1, \dots, x_m) = 0.    \label{QNE-intro}
\end{equation}
Furthermore, eliminating $\hat{z}_i$ from (\ref{QNE-intro}) and introducing $\hat{y} = \prod_{i=1}^m \hat{z}_i$, leads to the following operator relation
\be
\widehat{A}(\hat{x}_1,\ldots,\hat{x}_m,\hat y) P_C(x_1, \dots, x_m) = 0,
\ee
where $\widehat{A}(\hat{x}_1,\ldots,\hat{x}_m,\hat y)$ is a polynomial in $\hat{x}_i$ and $\hat{y}$ whose coefficients depend on $q$, and which can be interpreted as the quantum version of the resultant, i.e. it reduces to the resultant $A(x_1,\ldots,x_m,y)$ in the limit $q=1$.

In certain situations it is of interest to consider a single generating parameter $x$, by imposing that each $x_i$ variable -- or, more generally, a subset of $x_i$'s -- is proportional to a single $x$ (and treating proportionality factors, or remaining generating parameters, as constants or moduli). The asymptotics of (\ref{PC}) with such identifications can be encoded in an algebraic curve
\be
A(x,y) = 0,    \label{Axy}
\ee
which arises from the solution of the Nahm equations, or equivalently from an appropriate specialization of $x_i$'s in the resultant $A(x_1,\ldots,x_m,y)$. In what follows we refer to this curve as the (classical) quiver A-polynomial. Quiver A-polynomials have been already introduced and discussed in \cite{Panfil:2018sis,Ekholm:2018eee}. For quivers corresponding to knots, quiver A-polynomials are identified as various generalizations of A-polynomials for knots (hence the name ``quiver A-polynomials''). Similarly, for quivers arising in the reformulation of open topological string amplitudes, their A-polynomial are identified as mirror curves \cite{Panfil:2018faz}. However, apart from these special cases, it is of interest -- and the primary aim of this work -- to analyze A-polynomials associated to an arbitrary matrix $C$. In particular, interpreting the matrix $C$ as encoding a symmetric quiver, the corresponding A-polynomial captures certain classical (i.e. numerical) Donaldson-Thomas invariants or some combinations thereof, as also discussed in \cite{Garoufalidis:2015ewa,Panfil:2018sis}. 

Furthermore, there exists a quantum counterpart of (classical) quiver A-polynomial, i.e. an operator $\widehat{A}(\hat x, \hat y)$, which arises from the appropriate specialization of $x_i$'s in $\widehat{A}(\hat{x}_1,\ldots,\hat{x}_m,\hat y)$, and which encodes recursion relations for the series (\ref{PC}) (with the same specialization)
\be
\widehat{A}(\hat x, \hat y) P_C(x) = 0.
\ee
In this case $\hat y \hat x = q \hat x \hat y$. Similarly as in other contexts, we call this operator the quantum quiver A-polynomial. Its existence follows also from the fact that (\ref{PC}) is a generating series of $q$-holonomic functions (i.e. the summand in (\ref{PC}) is expressed in terms of $q$-Pochhammers and at most quadratic powers of $q$); it then follows from the theory of $q$-holonomic functions that it satisfies a $q$-difference equation, which can be written in terms of the operator $\widehat{A}(\hat x, \hat y)$ \cite{Garoufalidis-Le-survey}. In the limit $q=1$, quantum quiver A-polynomial reduces to the classical quiver A-polynomial. From quantum (quiver) A-polynomials one can also extract motivic Donaldson-Thomas invariants, or combinations thereof, or related BPS invariants arising upon an appropriate choice of the quiver, as discussed in \cite{Kucharski:2016rlb}. Nonetheless, explicit identification of a quantum A-polynomial for a given $C$ is usually a difficult task. 

\bigskip

There are two main goals of this work. The first one is to initiate a systematic analysis and classification of quiver A-polynomials. Their various properties have been reported in \cite{Panfil:2018sis}, however mainly in the context of other systems which were the main focus of that work. In this paper we are primarily interested in quiver A-polynomials in general, irrespective of their relation to knot theory or topological string theory. Among others, we discuss how to determine classical and quantum A-polynomials, and we identify various classes of A-polynomials of some specific genus (in particular genus 0).
 
Furthermore, the second important goal of this work is to show that the generating series (\ref{PC}) and corresponding quantum A-polynomials are encoded in the classical quiver A-polynomial, and this information can be extracted by means of the topological recursion, or equivalently the B-model formalism. 

The topological recursion is a framework which assigns an infinite family of multi-differential forms to a given algebraic curve, which is usually called a spectral curve in this context \cite{eyn-or}. Topological recursion was discovered in the context of matrix models, as a reformulation of loop equations \cite{Eynard:2004mh,Chekhov:2005rr,Chekhov:2006rq}. Subsequently it was shown that it has vast applications and enables to extract various quantities and invariants in problems, in which a certain algebraic curve plays a prominent role. For example, in the form of the remodeling conjecture it can be interpreted as the B-model formalism that enables to compute topological string amplitudes from the mirror curve \cite{BKMP}, it computes various enumerative invariants (e.g. various generalizations of Hurwitz numbers) \cite{Dumitrescu:2012dka}, expansions of colored knot polynomials \cite{DijkgraafFuji-2,Borot:2012cw}, etc. In \cite{abmodel,Bouchard:2016obz} it was shown in general how to determine quantum curves perturbatively by means of the topological recursion. In recent years various generalization of the topological recursion have been formulated, which include its refined version \cite{Chekhov:2010zg}, curves with higher ramification points \cite{Bouchard:2012yg}, formulation in terms of Airy structures \cite{Kontsevich:2017vdc,Andersen:2017vyk} and their supersymmetric generalization \cite{Bouchard:2019uhx}, etc.

More precisely, in this work we conjecture that the motivic generating series (\ref{PC}) can be interpreted as a wave function, which is annihilated by the quantum A-polynomial operator that arises as a quantization of a quiver A-polynomial. By means of the topological recursion we determine the expansion of (\ref{PC}) in $\hbar=\log q$ (so that $q=e^{\hbar}$) and confirm this statement in various examples. The quantization of classical A-polynomials that we consider follows general lines discussed in \cite{abmodel,Bouchard:2016obz}. However, note that in order to consider algebraic curves, we must consider specialization of various generating parameters $x_i$ to a single $x$, as explained above, so in general the topological recursion does not provide information about (\ref{PC}) with the full dependence on $x_i$, but about its specialized form that depends on a single $x$. In consequence, for a fixed specialization, the topological recursion encodes e.g. not all motivic Donaldson-Thomas invariants associated to the series (\ref{PC}), but some combinations thereof. Nonetheless, such specializations are also of interest in various applications (e.g. in the knots-quivers correspondence or the relation to topological string theory, as already mentioned above), so it is of importance to consider them. Furthermore, for a given $C$ one could conduct topological recursion calculations for different specializations of $x_i$, and then reconstruct more general information about (\ref{PC}).

Note that, even though the topological recursion is defined abstractly for spectral curves of arbitrary genus, in practice it can be computed for curves of genus 0 and 1. For this reason, it is difficult, or just impossible, to test it explicitly in many situations of interest, for example for knots whose A-polynomials are of genus larger than 1, or for toric Calabi-Yau manifolds whose mirror curves have genus larger than 1; in those cases the number of interesting, directly computable examples is limited. Happily, for $q$-series of the form (\ref{PC}) we have much larger testing ground -- there are many matrices $C$, of various size, for which corresponding A-polynomials are of genus 0 or 1, and thus our conjecture can be explicitly verified. In this work we conduct such tests and confirm the above conjecture for a number of representative quivers whose A-polynomials have genus 0  (apart form one subtlety for a class of ``diagonal'' quivers).  

Furthermore, also note that in some special cases our conjecture follows from combining various earlier results. For example, it is shown in \cite{Panfil:2018faz} that open topological string amplitudes for toric ``strip'' geometries take form of motivic generating functions for certain quivers. On the other hand, according to the remodeling conjecture \cite{BKMP} these amplitudes are reproduced by the topological recursion for mirror curves, and it follows from \cite{Panfil:2018faz} that in this case mirror curves can be identified with quiver A-polynomials. A prototype example in this case are topological string amplitudes and the quantum curve for (framed) $\mathbb{C}^3$, for which the topological recursion formalism is discussed e.g. in \cite{abmodel,vb-ps}, and for which the corresponding quiver consists of one vertex with a number of loops. Combining these facts we conclude, that the topological recursion for mirror curves for strip geometries reproduces motivic generating functions for quivers corresponding to such geometries. The results of this paper can be regarded as a generalization of this observation to much larger family of $q$-series (\ref{PC}), whose structure is encoded in arbitrary matrices $C$. This also means that various quantities associated to those $q$-series -- such as Donaldson-Thomas invariants, various knot invariants that arise via the knots-quivers correspondence upon an appropriate choice of $C$, certain amplitudes in integrable theories, etc. -- have the structure compatible with the topological recursion, and can be computed using this recursion. We believe this is an important conceptual statement, which should be of advantage in further analysis of the above mentioned quantities, and which shows yet another powerful application of the topological recursion. 

In particular, we stress that some of our results explicitly illustrate the conjecture that colored knot invariants are reconstructed by the topological recursion. For example, the generating function ({\ref{PC}) and A-polynomials for the quiver which has one vertex and $f$ loops (the same one as mentioned above, encoding topological string amplitudes for $\mathbb{C}^3$), capture extremal invariants of the unknot in framing $f$. On the other hand, the quiver ${2\ 1 \brack 1\ 1}$ encodes full colored HOMFLY polynomials for the unknot in framing $f=1$ \cite{Kucharski:2017ogk}. More complicated examples involve quivers ${2\ 0 \brack 0\ 0}$ and ${2\ 1 \brack 1\ 0}$ (and their framed relatives), whose A-polynomials also have genus 0, and which capture colored extremal invariants of left-handed and right-handed trefoil knot, and (in some specific framing) also encode counting of certain Duchon paths \cite{Panfil:2018sis}. In what follows, among others, we discuss properties of these particular quivers, and show how they are captured by the topological recursion. This supports the conjecture relating knot invariants and topological recursion \cite{DijkgraafFuji-2,Borot:2012cw,abmodel}, and also illustrates new links between path counting problems and topological recursion. We conjecture that analogous relations to the topological recursion hold for other other types of Duchon paths, at least those which are related (as discussed in \cite{Panfil:2018sis}) to other quivers.

The plan of this paper is as follows. In section \ref{sec-A} we introduce classical and quantum A-polynomials for quivers and summarize their general properties. In section \ref{sec-our-quivers} we initiate systematic classification of quiver A-polynomials; among others we identify various families of quivers of some specific genus (in particular genus 0) and derive their classical and quantum A-polynomials. In section \ref{sec-tr-intro} we concisely review the formalism of the topological recursion and its relations to quantum curves. In section \ref{sec-TR} we conduct explicit tests of our conjecture, and confirm (apart from one subtlety for ``diagonal'' quivers mentioned above) that (specializations of) the series (\ref{PC}) and corresponding quantum curves, associated to various representative quivers identified in section \ref{sec-our-quivers}, are determined by the topological recursion.

%***************************************************************************************************
%***************************************************************************************************

%\section{Nahm sums, quiver generating series, and A-polynomials}

\section{Classical and quantum quiver A-polynomials}  \label{sec-A}

One of the main characters in this work is the following series
\begin{equation}
  P_C(x_1, \dots, x_m) = \sum_{d_1, \dots, d_m\geq 0}\frac{(-q^{1/2})^{\sum_{i,j=1}^m C_{i,j} d_i d_j}}{(q; q)_{d_1}\cdots (q;q)_{d_m}} x_1^{d_1} \cdots x_m^{d_m},    \label{QGF}
\end{equation}
where $C_{i,j}$ are entries of a fixed symmetric matrix $C$ of size $m$, and $x_1,\ldots, x_m$ are interpreted as generating parameters, and the $q$-Pochhammer symbol is $(x;q)_d = \prod_{i=0}^{d-1}(1-xq^i)$. We also introduce the following notation for coefficients in this series
\begin{align}
\begin{split}
P_C(x_1, \dots, x_m) &= \sum_{d_1, \dots, d_m \geq 0}    P_{d_1, \dots, d_m} x_1^{d_1} \cdots x_m^{d_m}, \\
P_{d_1, \dots, d_m} &= \frac{(-q^{1/2})^{\sum_{i,j=1}^m C_{i,j} d_i d_j}}{(q; q)_{d_1}\cdots (q;q)_{d_m}}.   \label{Coeffs}
\end{split}
\end{align}
Furthermore, for $q=e^{\hbar}$, such series has the following asymptotic expansion
\be
P_C(x_1, \dots, x_m) = \exp\Big(\frac{1}{\hbar}S_0(x) + S_1(x) + \hbar S_2(x) + \ldots \Big).   \label{PC-Sk}
\ee 

As discussed in the introduction, such series play an important role in physics and mathematics. On one hand such series are referred to as Nahm sums, and for specific values of $C$ and $x_i$ arise as characters of integrable theories. On the other hand, they arise as motivic generating series associated to a symmetric quiver determined by the matrix $C$ (so that $C_{i,j}=C_{j,i}$ is the number of arrows from vertex $i$ to vertex $j$); a product decomposition of such a series into quantum dilogarithms encodes motivic Donaldson-Thomas invariants for a quiver in question. For specific choices of $C$ and identification of $x_i$ with a single parameter $x$, such series encode knot invariants and topological string amplitudes. In this section we discuss how to derive classical and quantum quiver A-polynomials associated to series (\ref{QGF}) and summarize some of their properties.

%***************************************************************************************************

\subsection{Classical quiver A-polynomials}   \label{ssec-classical-A}

An interesting information is encoded in the asymptotics of the series (\ref{QGF}), in the limit $\hbar\to 0$. It can be obtained by replacing the sums over $d_i$ by integrals over $z_i = e^{\hbar d_i}$
\begin{equation}
P_C(x_1, \dots, x_n) \simeq \int%_{1}^{\infty} 
\frac{dz_1\cdots dz_m}{ z_1 \cdots z_m} \exp\Big(\frac{1}{\hbar} S_0(x,z)+\dots\Big),   \label{P-C-int}
\end{equation}
with the leading term
\begin{equation}
S_0(x,z) = \frac{1}{2}\sum_{i,j=1}^m C_{i,j}\log z_i  \log z_j + \sum_{i=1}^m\Big( \log z_i \log x_i + {\rm Li}_2(z_i) - {\rm Li}_2(1) + i \pi  C_{i,i} \log z_i \Big),     \label{S0-quiver}
\end{equation} 
whose form follows from the relation $(x;q)_{d} \simeq e^{ \frac{1}{\hbar}({\rm Li}_2 (x) - {\rm Li_2}(q^d x)) + \ldots}$, and where the dilogarithm function is ${\rm Li}_2(x)=\sum_{i=1}^{\infty} \frac{x^i}{i^2}$. Exponentiating the equation for stationary points $\partial_{z_i} S_0(x,z)= 0$ leads to a set of Nahm equations
\begin{equation}  
H_i \equiv  (-1)^{C_{i,i}}x_i \prod_{j=1}^m z_j^{C_{i,j}} + z_i -1 = 0, \qquad i=1,\ldots, m.  \label{clNahm}
\end{equation}
We denote the solution of these equations by $\bar{z}=(\bar{z}_i)_{i=1,\ldots,m}$, and we also introduce
\be
y_i = y_i(x_1,\ldots,x_m) = e^{x_i \partial_{x_i} S_0(x,\bar{z})} = \bar{z}_i (x_1,\ldots,x_m),   \label{yi-def}
\ee
as well as 
\be
y(x_1,\ldots,x_m) = e^{\sum_{i=1}^m \partial_{x_i} S_0(x,\bar{z})} = \prod_{i=1}^m y_i(x_1,\ldots,x_m) = \prod_{i=1}^m \bar{z}_i.    \label{y-def}
\ee

The crucial role in this work is played by the quiver A-polynomial, whose form follows from the above equations. One way to introduce it is to consider first the following mixed resultant 
\begin{equation}
A(x_1,\dots,x_m,y) = \mathrm{res}_{z_1,\dots,z_m} \Big\{y-\prod_{j=1}^m z_j = 0,\ H_1 = 0,\ldots,\ H_m = 0\Big\}.    \label{resultant}
\end{equation}
For a proper definition of the mixed resultant see \cite{GKZ}; essentially, given a collection of Laurent polynomials $f_0,\dots,f_n$, its mixed resultant is the unique (up to a sign) irreducible polynomial in coefficients of $f_i$, which vanishes if and only if $f_0,\dots,f_n$ have a common root.  The mixed resultant, as a polynomial in $y$ and $x_i$, has integer coefficients, it is irreducible over integers, and it is tempered. In various situations it is of interest to identify all (or some) $x_i$ in the quiver generating function $P_C(x_1,\ldots,x_m)$ and in $A(x_1,\dots,x_m,y)$ with a single variable $x$ (possibly up to some proportionality factors). For example, in the knots-quivers correspondence, all $x_i$'s are proportional to such a single $x$ \cite{Kucharski:2017poe,Kucharski:2017ogk}, while in the relation to topological strings on strip geometries only $x_1$ is identified with $x$ (and other $x_i$'s are identified with K{\"a}hler moduli) \cite{Panfil:2018faz}. Upon such an identification, we denote by $P_C(x)$ the quiver generating function with identified parameters, while the above resultant reduces to a polynomial in $x$ and $y$ (possibly depending on some extra parameters), which we refer to as the quiver A-polynomial, and whose zeros define a complex curve in $\mathbb{C}^*\times\mathbb{C}^*$ 
\be
A(x,y) = 0.    \label{Axy}
\ee
This equation can be also obtained as a direct solution of the set of equations (\ref{clNahm}) and (\ref{y-def}), with appropriate identification of $x_i$'s. Moreover, the leading term $S_0$ in the expansion (\ref{PC-Sk}), when various $x_i$ are identified with $x$, is given by
\be
S_0 = \int \log y \frac{dx}{x},   \label{PC-S0}
\ee
where $y=y(x)$ is the solution of the equation (\ref{Axy}). This solution can be equivalently obtained as the limit
\be
y=y(x) = \lim_{q\to 1} \frac{P_C(qx)}{P_C(x)}.    \label{y-limit-PC}
\ee
Higher order terms $S_k(x)$ in (\ref{PC-Sk}) can be determined by WKB method.

One of the main aims of this paper is the analysis of such quiver A-polynomials.

%***************************************************************************************************

\subsection{Quantum quiver A-polynomials}  \label{ssec-quantum-A}

Information about the generating series (\ref{QGF}) is equivalently encoded in recursion relations it satisfies. At the same time, these recursion relations can be thought of as quantum generalizations of classical equations from the previous section. Such recursion relations can be determined in a number of ways, as we summarize in this section. In the last section of this paper we show, how such a quantization arises from the formalism of topological recursion. 

One way to determine recursion relations satisfied by (\ref{QGF}) is to consider a shift of one of the subscripts in (\ref{Coeffs}). Such a shift can be written as
\be
P_{d_1, \dots,d_k+1 ,\dots, d_m} =  \frac{(-q^{1/2})^{\sum_{i,j=1}^m C_{ij} (d_i+\delta_{ik}) (d_j+\delta_{jk})}}{(q; q)_{d_1} \cdots (q;q)_{d_k+1} \cdots (q;q)_{d_m}} 
     =  \frac{(-q^{1/2})^{C_{kk} + 2 \sum_i^m C_{ki} d_i}}{1-q^{d_k+1}} P_{d_1, \dots, d_m}.
\ee
Introducing operators $\hat{z}_l$ such that 
\be
\hat{z}_l \hat{x}_k = q^{\delta_{kl}} \hat{x}_k \hat{z}_l,
\ee
we can write difference equations for the generating series of the above coefficients as follows
\be
\sum_{(d_1, \dots, d_m)\geq 0} P_{d_1, \dots,d_k+1 ,\dots, d_m} \prod_i^m x_i^{d_i+\delta_{ki}} = \Big( (1-\hat{z}_k)^{-1}(-q^{1/2})^{C_{kk}} \hat{x}_k \prod_j^m \hat{z}_j^{C_{kj}} \Big) \cdot P(x_1, \dots, x_m) .
\ee
Multiplying both sides by $(1-\hat{z}_k)$ leads to the following set of equations
\begin{equation}
    \hat{H}_k  P(x_1, \dots, x_m) = 0, \quad \textrm{for}\ \hat{H}_k = (-q^{1/2})^{C_{kk}} \hat{x}_k \prod_j^m \hat{z}_j^{C_{kj}} + \hat{z}_k - 1.    \label{QNE}
\end{equation}
These are quantum counterparts of the classical equations (\ref{clNahm}). 

Furthermore, analogously as in the classical case, one can identify various $x_i$ with a single variable $x$. It is then natural to introduce the corresponding shift operator; in particular, for all $x_i=x$, the shift operator takes form
\be
\hat y = \prod_{k=1}^m \hat{z}_k,     \label{y-hat-z-hat}
\ee
and it satisfies $\hat y \hat x=q\hat x \hat y$. One can then eliminate appropriate $\hat{z}_k$'s from (\ref{QNE}), which leads to a single difference equation for the generating function (\ref{QGF}) 
\be
\widehat{A}(\hat x_1,\ldots,\hat x_m,\hat y) P_C(x_1,\ldots,x_m) = 0,    \label{Ahat-all-x-Px}
\ee
which is a quantum counterpart of the resultant (\ref{resultant}). Furthermore, appropriately identifying $x_k$'s (e.g. as $x_i=c_i x$, or simply $x_i=x$), so that a single variable $x$ remains, we obtain a difference equation of the form
\be
\widehat{A}(\hat x,\hat y) P_C(x) = 0.    \label{AhatPx}
\ee
We refer to $\widehat{A}(\hat x,\hat y)$ as the quantum quiver A-polynomial. The above equation can be also expanded in $\hbar$, leading to a set of differential equations for coefficients $S_k(x)$ introduced in (\ref{PC-Sk}). On the other hand, in the classical $q\to 1$ limit, $\widehat{A}(\hat x,\hat y)$ reduces to the classical quiver A-polynomial (\ref{Axy}), which is determined just by $S_0(x)$.

Another strategy to determine quantum quiver A-polynomial is to take advantage of computer programs for symbolic computations, such as qZeil and related ones \cite{qZeil}. For brevity, let us consider a quiver with 2 vertices, and suppose that we identify the generating parameters as $x_1=x_2=x$. The generating function (\ref{QGF}) in this case can be written as 
\be
P_C(x) = \sum_{r=0}^{\infty} P_r x^r,
\ee
where
\begin{align}
\begin{split}
P_r &= \sum_{d_1 + d_2 = r}\frac{(-q^{1/2})^{\sum_{i,j=1}^2 C_{i,j} d_i d_j}}{(q; q)_{d_1} (q;q)_{d_2}} = \\
&= \sum_{d_1 = 0}^r (-1)^{C_{1,1} d_1 + C_{2,2} d_2 } \frac{  q^{\frac12 (  C_{1,1} d_1^2 + 2C_{1,2} d_1(r-d_1) + C_{2,2} (r-d_1)^2 )  }}{(q; q)_{d_1} (q;q)_{r-d_1}}.
\end{split}
\end{align}
Importantly, a recursion relation satisfied by $P_C(x)$ is equivalent to a recursion relation satisfied by $P_r$. Moreover, $P_r$ given above are $q$-holonomic functions, so it is guaranteed that they satisfy recursion relations of finite order in operators $\widehat{M}$ and $\widehat{L}$, which act as
\be
\widehat{M} P_r = q^r P_r,\qquad \widehat{L}P_r= P_{r+1},
\ee
and which satisfy the relation $\widehat{L}\widehat{M} = q \widehat{M}\widehat{L}$. Such recursion relations for $P_r$ can be found e.g. by the above metioned qZeil environment \cite{qZeil}. In what follows we will determine such recursions for various matrices $C$ using this tool. Furthermore, the operators $\widehat{M}$ and $\widehat{L}$ are dual to $\hat x$ and $\hat y$, i.e.
\begin{align}
\begin{split}
\widehat{M} P_C(x) &= \sum_{r=0}^{\infty} P_r (qx)^r = P_C(qx), \\
\widehat{L} P_C(x) & = \sum_{r=0}^{\infty} P_{r+1} x^r = \frac{1}{x}\big( P_C(x) -P_0  \big).  \label{MP-LP}
\end{split}
\end{align}
Therefore, if $\widetilde{A}(\widehat{M}, \widehat{L})$ is an operator that encodes recursion relations for $P_r$, i.e. $\widetilde{A}(\widehat{M}, \widehat{L}) P_r = 0$, then the recursion relations for the generating series $P_C(x)$ are captured by the operator
\be
\widehat{A}(\hat x, \hat y) = \widetilde{A}(\hat y,\hat{x}^{-1}),   \label{Ahat-dual-2vertex}
\ee
see also \cite{Garoufalidis:2015ewa}. However, because of the shift in the action of $\widehat{L}$ in (\ref{MP-LP}), the recursion for $P(x)$ takes a nice form 
\be
\widehat{A}(\hat x, \hat y) P_C(x)=0      \label{Ahat-P_C2vertex}
\ee
only if boundary conditions related to this shift get compensated. If this is not the case, the recursion may get modified by some extra terms, which can be detected explicitly by verifying the result of $A(\hat x, \hat y) P(x)$ for first few terms in $x$-expansion of $P(x)$. In case of quiver generating functions, we have verified for a large set of matrices $C$ of size 2, that such boundary terms indeed vanish and the equation (\ref{Ahat-P_C2vertex}) holds without any extra terms.

We also state some basic properties of quantum A-polynomials. First, if $\psi(x)$ is annihilated by quantum A-polynomial, $\hat{A}(\hat{x}, \hat{y}) \psi(x) = 0$, then 
\be
 \hat{A}(q^b \hat{x}, q^{-a} \hat{y}) \widetilde{\psi}(x) = 0,\qquad \textrm{for}\quad \widetilde{\psi}(x) = x^a \psi(q^b x).   \label{rescale-Ahat}
 \ee
Second, rescaling the argument of the wave-function $\psi(x)=\exp(\hbar^{-1}S_0(x) + S_1(x) +\ldots)$ (which is annihilated by $\hat{A}(\hat{x}, \hat{y})$) by $q$, results in shifting $S_1(x)$ by $\log y$:
\be
\psi(qx) = \exp\Big(\frac{1}{\hbar}S_0(x) + \big(S_1(x) + \log y\big) + \ldots \Big),   \label{psi-qx}
\ee
and of course also in appropriate shifting of subsequent $S_k$ with $k>1$. 

%***************************************************************************************************

\subsection{Framing and reciprocity}

In what follows we consider various operations on the generating function (\ref{QGF}), which do not change genus of corresponding A-polynomials. The first such operation is a shift of each element of the matrix $C$ by the same number
\be
C \mapsto C_f = C + \left[\begin{array}{ccc} f & f & \cdots\\
f & f & \cdots \\ 
\vdots & \vdots & \ddots\end{array} \right]   \label{framing}
\ee
We refer to this operation as (a change of) \emph{framing}, because it indeed corresponds to a change of framing in case the generating function (\ref{QGF}) for appropriately chosen $C$ encodes knot polynomials or topological string amplitudes \cite{Panfil:2018sis}. Note that the operation (\ref{framing}) modifies the summand in (\ref{QGF}) by 
\be
(-q^{1/2})^{f \sum_{i,j} d_i d_j } = (-q^{1/2})^{f(d_1+\ldots d_m)^2} = (-q^{1/2})^{f r^2}
\ee
for $r=d_1+\ldots + d_m$. In particular, if in the series (\ref{QGF}) we identify $x_i=\pm x$ and write this series as $P_C(x) = \sum_r P_r x^r$, then the change of framing (\ref{framing}) is equivalent to multiplying the coefficient $P_r$ by $(-q^{1/2})^{f r^2}$. It also follows that if such $P_C(x)$ satisfies a difference equation $\widehat{A}(\hat x, \hat y)P_C(x)=0$, then the framed generating function $P_{C_f}(x)$ satisfies a difference equation with $\hat x$ replaced by $\hat x (-\hat y)^f$, i.e.
\be
\widehat{A}(\hat x(- \hat y)^f,\hat y) P_{C_f}(x) = 0.
\ee
Clearly, after the framing operation, the corresponding classical curve takes form $A(x(-y)^f,y)=0$, where $A(x,y)=0$ is the curve corresponding to the original generating function. An important statement is that such an operation does not change the genus of the algebraic curve. In particular, for curves of genus 0 that have a rational parametrization $(x(t),y(t))$, the framed curve has also a rational parametrization $(x(t)(-y(t))^f,y(t))$ and thus its genus is also 0.

Another operation that we also consider is 
\be
C\mapsto 1_{m\times m}-C = \left[\begin{array}{ccc} 1 & 0 & \cdots\\
0 & 1 & \cdots \\ 
\vdots & \vdots & \ddots\end{array} \right]  - C,   \label{reciprocal}
\ee
where $1_{m\times m}\equiv 1$ is the identity matrix of size $m$. We call the quiver characterized by the resulting matrix $1-C$ as the reciprocal quiver. In particular, in the knots-quivers correspondence, for a matrix $C$ corresponding to some knot, the matrix $1-C$ corresponds to the mirror image of this knot. Note that the form of quantum and classical A-polynomial associated to the matrix $(1-C)$ can be deduced from A-polynomials associated to $C$. To show that, we first denote by $\widehat{H}^{1-C}_i\equiv \widehat{H}^{1-C}_i(x_j,\hat z_j,q)$ the operator (\ref{QNE}) associated to the matrix $(1-C)$, and rewrite it as
\begin{align}
\begin{split}
     \widehat{H}^{1-C}_i(x_j,\hat{z}_j,q) & =  (-q^{1/2})^{\delta_{ii}-C_{ii}} \hat{x}_i \prod_j^m \hat{z}_j^{\delta_{ij}-C_{ij}} + \hat{z}_i - 1 = \\
 %     & =  (-q^{-1/2})^{C_{ii}} (-q^{1/2}) \hat{x}_i \hat{z}_i \prod_j^m \hat{z}_j^{-C_{ij}} + \hat{z}_i - 1  \\
      & =  (-\hat{z}_i) \Big((-q^{-1/2})^{C_{ii}} (q^{-1/2} \hat{x}_i) \prod_j^m \hat{z}_j^{-C_{ij}} - 1 + \hat{z}_i^{-1}\Big) = \\
       & =  (-\hat{z}_i) \widehat{H}^C_i(q^{-1/2} x_j,\hat{z}_j^{-1},q^{-1}),  \label{RecQNE}
\end{split}
\end{align}
where $\widehat{H}^C_i(q^{-1/2} x_j,\hat{z}_j^{-1},q^{-1})$ is the operator (\ref{QNE}) associated to $C$, but with modified arguments. Furthermore, we can rewrite the defining equation of $\hat y$ in (\ref{y-hat-z-hat}), acting on the generating function $P_{1-C}(x_1,\ldots,x_m)$ associated to $(1-C)$, as follows
\begin{align}
\begin{split}
    0&=(\hat{y} - \prod_{j=1}^m \hat{z}_j^{\alpha_j}) \cdot P_{1-C}(x_1,\ldots,x_m) = \\
    &= \Big(-\hat{y} \prod_{j=1}^m \hat{z}_j^{\alpha_j}\Big) \Big(\hat{y}^{-1} - \prod_{j=1}^m \hat{z}_j^{-\alpha_j}\Big)  P_{1-C}(x_1,\ldots,x_m).   \label{RecQNE2}
\end{split}
\end{align}
The operator (\ref{RecQNE}) and the second line of (\ref{RecQNE2}) (ignoring an irrelevant first factor $(-\hat{y} \prod_j \hat{z}_j^{\alpha_j})$) are analogous to the pair of equations (\ref{QNE}) and (\ref{y-hat-z-hat}), and (via modification of arguments) relate operators associated to the matrix $(1-C)$ and those associated to $C$. Eliminating $\hat{z}_j^{-1}$ from the above equations implies, that if $\widehat{A}_C(x_j,\hat y,q)$ is the operator that annihilates the generating function (\ref{QGF}) associated to $C$, then the same operator but with modified arguments is associated to $1-C$
\be
\widehat{A}_{1-C}(x_j,\hat{y},q) = \widehat{A}_C(q^{-1/2}x_j,\hat{y}^{-1},q^{-1}),
\ee
and therefore the above operator annihilates the generating function $P_{1-C}(x_1,\ldots,x_m)$ 
\be
\widehat{A}_C(q^{-1/2}x_j,\hat{y}^{-1},q^{-1}) P_{1-C}(x_1,\ldots,x_m) = 0.
\ee
In consequence, analogous relation between quantum A-polynomials holds after identifying $x_j$ with a single $x$, $\widehat{A}_{1-C}(x,\hat{y},q) = \widehat{A}_C(q^{-1/2}x,\hat{y}^{-1},q^{-1})$, while classical resultant and A-polynomial associated to $1-C$ are related to those associated to $C$ by inverting $y$
\be
A_{1-C}(x,y) = A_C(x,y^{-1}).
\ee

%***************************************************************************************************
%***************************************************************************************************

\section{Classification and examples of quiver A-polynomials}    \label{sec-our-quivers}

A basic characteristic of an algebraic curve is its genus. In particular it plays a crucial role from the perspective of the topological recursion -- in practice, explicit topological recursion calculations can be conducted for spectral curves of genus 0 or 1. It is therefore of interest to determine the genus a quiver A-polynomial associated to a matrix $C$, and identify quiver A-polynomials whose genus is 0 or 1. In this section we make the first step towards such a classification. After a quick summary in section \ref{ssec-1vertex} of the simplest case of $C$ of size 1, in section \ref{ssec-2vertex} we consider arbitrary symmetric matrices $C$ of size 2, determine genus of associated A-polynomials, and identify and present in more detail all families of A-polynomials of genus 0. Finally, in sections \ref{ssec-uniform} and \ref{ssec-family} we identify two families of matrices $C$ of arbitrary size, whose A-polynomials have genus 0. Note that when discussing particular examples we focus on analysis of A-polynomials that we call admissible, i.e. they are irreducible and have a maximal number of ramification points (which cannot be increased by changing the framing). In section \ref{sec-TR} we will show that the topological recursion reproduces quiver generating series and corresponding quantum curves for those examples.
 
%Let us present some examples of quiver A-polynomials. We present primarily examples of curves of genus zero, as they provide a nice playground for the topological recursion, which is of our main interest in this paper. One of the prominent features of algebraic curves of genus zero is that they have a rational parametrization, which we also provide in some examples; note that a parametrization of a spectral curve is an essential ingredient of the topological recursion. Furthermore, to take advantage of the topological recursion, we focus on analysis of curves that we call admissible, i.e. they are irreducible and have a maximal number of ramification points (which cannot be increased by changing the framing). In section \ref{sec-TR} we will show that the topological recursion reproduces quiver generating series and corresponding quantum curves, via analysis of particular examples of quivers from the families presented below.

%***************************************************************************************************

\subsection{One-vertex quiver, $m=1$}  \label{ssec-1vertex}

We start the analysis from quivers that consist of one (i.e. $m=1$) vertex and $f$ loops, so that the quiver matrix reads $C=[f]$ with $f\in\mathbb{Z}$. The quiver generating function (\ref{QGF}) in this case takes form
\begin{equation}
  P_C(x) = \sum_{d \geq 0}\frac{(-q^{1/2})^{f d^2}}{(q; q)_{d}} x^{d} = \sum_{d=0}^{\infty} (-1)^{f d} \frac{q^{\frac{f}{2}d^2}}{(q,q)_d}x^d \equiv \sum_{d=0}^{\infty} p_d x^d,    \label{QGF-1vertex}
\end{equation}
and it can be regarded as the framed version (\ref{framing}) of the $C=[0]$ case. Note that
\be
\frac{p_d}{p_{d-1}} = \frac{(-1)^f q^{\frac{f}{2}(2d-1)}}{1-q^d}  \qquad \Rightarrow \qquad
p_d - q^d p_d - (-1)^f q^{\frac{f}{2}(2d-1)} p_{d-1} = 0,
\ee
so multiplying the last equation by $x^d$ and summing over $d$ we get the expression which can be written as $\widehat{A}(\hat{x},\hat{y}) P_C(x)\sim0$, where $\sim$ means equality possibly up to some initial terms, and where the difference operator has form
\be
\widehat{A}(\hat{x},\hat{y}) = (-1)^f q^{f/2} \hat{x} \hat{y}^f +\hat{y} - 1.   \label{Ahat-1vertex}
\ee
We verify that the quantum A-polynomial is satisfied without any extra terms, so that
\be
\widehat{A}(\hat{x},\hat{y}) P_C(x) = 0.
\ee
On the other hand, the saddle point approximation of $P_C(x)$ yields
\begin{align}
\begin{split}
	S_0(x) &= -{\rm Li}_2(1-y) - \frac{f}{2} (\log y)^2, \\
	S_1(x) &= - \frac{1}{2}\log \frac{(1-y)y}{x} - \frac{1}{2} \log\left( - \frac{dx}{dy}\right),  \label{S0S1-1vertex}
\end{split}
\end{align}
where $y(x)$ is a solution to the classical A-polynomial 
\be
A(x,y) = x(-y)^f + y - 1 = 0,  \label{SPE}
\ee
which is indeed $q=1$ limit of (\ref{Ahat-1vertex}). For each $f$ this curve has the following rational parametrization
\begin{equation}
	x(t) = (-1)^f \frac{1-t}{t^f}, \qquad y(t) = t,    \label{1vertex-param}
\end{equation}
which immediately implies that it has genus 0. Moreover, solving the equation (\ref{SPE}) for $y=y(x)$ we find 
\begin{align}
\begin{split}
	y(x) =&\, 1 - (-1)^f x + f x^2 - (-1)^f \frac{3f - 1}{2} x^3 + \frac{8f^2 - 6f + 1}{3} x^4 + \\
	&- (-1)^f \frac{f (125 f^3 - 150 f^2 + 55f - 6)}{24}  x^5 + \mathcal{O}(x^6).
\end{split}
\end{align}
We recall that the generating function (\ref{QGF-1vertex}) also arises as the generating function of extremal HOMFLY polynomials for the unknot \cite{Kucharski:2017ogk}, and as the partition function for a brane in $\mathbb{C}^3$ \cite{Panfil:2018sis}.

%In this case the leading term (\ref{S0-quiver}) takes form
%\begin{equation}
%  S_{0}(z) =  {\rm Li}_2(z) + \log z \log((-1)^f x) + \frac{1}{2}f(\log(z))^2 - \frac{\pi^2}{6}, %\label{single_action}
%\end{equation}
%which leads to the following saddle point equation, which at the same time can be identified with %quiver A-polynomial
%\begin{equation} \label{SPE}
%A(x,y) = (-1)^f xy^f+y-1=0. 
%%, \qquad x>0, \qquad 0<z<1, \qquad c\in \mathbf{N}
%\end{equation}

%***************************************************************************************************

\subsection{Two-vertex quivers, $m=2$}   \label{ssec-2vertex}

Now we consider symmetric matrices $C$ of size 2, with the aim of determining the genus of their A-polynomials, and in particular identifying all A-polynomials of genus 0. These matrices depend on 3 parameters; however, recalling that the framing operation (\ref{framing}) does not change the genus, we can use it to remove off-diagonal terms. Thus it is sufficient to focus on diagonal matrices
\be
\left[\begin{array}{cc}
a & 0 \\
0 & b \\
\end{array}\right]    \label{diagonal-ab}
\ee
Moreover, let us first consider the specialization of generating parameters $x_1=c_1x$ and $x_2=c_2 x$ with generic $c_1,c_2\in\mathbb{C}$, and look for A-polynomials which do not factorize (in what follows we are primarily interested in specialization $c_1=\pm c_2=1$, and then for some $C$ the A-polynomial may be reducible). We determine such A-polynomials along the lines presented in section \ref{ssec-classical-A}, or equivalently as $q=1$ limit of quantum A-polynomials, determined as discussed in section \ref{ssec-quantum-A}. From this analysis we find, first of all, that A-polynomials for a fixed genus $g>0$ arise for a finite number of pairs $(a,b)$, as summarized below. 

On the other hand, we find a few infinite families of A-polynomials of genus 0, which arise for pairs $(a,b)$ of the form $(a,0), (a,1), (a,a)$ and $(a,1-a)$, for $a\in\mathbb{Z}$. In fact, the pairs $(a,0)$ and $(a,1)$ are related by the reciprocity operation (\ref{reciprocal}) that does not change the genus of A-polynomial; however it is of advanatage to write down explicitly results for all those cases. For all such pairs $(a,b)$, the corresponding A-polynomials factorize only when $x_1=x_2$ (i.e. $c_1=c_2=1$) for the family $(a,1-a)$, and for $(a,a)$ for $a\neq 0,1$; in all other cases A-polynomials are irreducible (for all $c_1$ and $c_2$). Let us now provide explicit results for quantum and classical curves, as well as their parametrizations, for all these families that yield curves of genus 0. 

\subsubsection*{Trivial quiver ${0\ 0 \brack 0\ 0}$}

To start with, consider the simplest uniform case
\begin{equation}C = 
    \begin{bmatrix}
    0 & 0 \\
    0 & 0
    \end{bmatrix}
\end{equation}
The operators that annihilates the generating function (\ref{QGF}), and its specializations $(x_1 = c_1 x,x_2 = c_2 x)$ and $x_1=x=\pm x_2$, take form
\begin{align}
\begin{split}
    \hat{A}(x_1,x_2,\hat{y}) &=  (1 - x_1) (1 - x_2) - \hat{y}, \\
    \hat{A}(x_1=c_1x,x_2=c_2x,\hat{y}) &=   (1 - x c_1) (1 - x c_2) - \hat{y}, \\
    \hat{A}(x_1=x,x_2=x,\hat{y}) &=  (1 - x)^2 - \hat{y}, \\
    \hat{A}(x_1=x,x_2=-x,\hat{y}) &= (1 - x) (1 + x) - \hat{y}.
\end{split}
\end{align}
The resultant and corresponding classical A-polynomials take form
\begin{align}
\begin{split}
    A(x_1,x_2,y) &=  (1 - x_1) (1 - x_2) - y, \\
    A(x_1=c_1x,x_2=c_2x,y) &= (1 - x c_1) (1 - x c_2) - y, \\
    A(x_1=x,x_2=x,y) &=  (1 - x) (1 - x) - y, \\
    A(x_1=x,x_2=-x,y) &= (1 - x) (1 + x) - y,
\end{split}
\end{align}
and their parametrizations read
\begin{align}
\begin{split}
    x(t) = t, & \quad y(t) = (1 - c_1 t) (1 - c_2 t),  \\
    x(t) = t, & \quad y(t) = (1 - t)^2, \qquad \textrm{for}\   c_1 = c_2 = 1, \label{ParamUni11}\\
    x(t) = t, & \quad y(t) = (1 - t) (1 + t), \qquad \textrm{for}\  c_1 = - c_2 = 1 .
\end{split}
\end{align}
%Note: These are not the parametrization in which we preformed the computations in TR section file. We did the computation only for $c_1=c_2=1$. The parametrization \eqref{ParamUni11} relates to that of TR section by $t \rightarrow 1 - t$.

\subsubsection*{Family ${a\ 0 \brack 0\ 0}$}

The above uniform quiver is a special example of the more general family of the form
\begin{equation}
    C=\begin{bmatrix}
    a & 0 \\
    0 & 0
    \end{bmatrix}    \label{family-a0}
\end{equation}
for $a\in \mathbb{Z}$. Recall that the data associated to such a quiver with $a=1$ encodes colored HOMFLY polynomials for the unknot \cite{Kucharski:2017ogk}, while for $a=2$ it captures extremal invariants of the trefoil knot and certain Duchon paths \cite{Panfil:2018sis}. 

In general, the form of quantum A-polynomial in this case depends on whether $a$ is positive or not. For $a\geq 1$ we find
\begin{align}
\begin{split}
    \hat{A}(x_1,x_2,\hat{y}) &= (-q^{1/2})^a x_1 \hat{y}^a + (q x_2;q)_{a - 1} \hat{y} - (x_2;q)_a, \\
    \hat{A}(x_1=c_1x,x_2=c_2x,\hat{y}) &=  (-q^{1/2})^a x c_1 \hat{y}^a + (q x c_2;q)_{a - 1} \hat{y} - (x c_2;q)_a, \\
    \hat{A}(x_1=x,x_2=x,\hat{y}) &=  (-q^{1/2})^a x \hat{y}^a + (q x;q)_{a - 1} \hat{y} - (x;q)_a, \\
    \hat{A}(x_1=x,x_2=-x,\hat{y}) &=  (-q^{1/2})^a x \hat{y}^a + (-q x;q)_{a - 1} \hat{y} - (-x;q)_a, \\
\end{split}
\end{align}
where $(x,q)_a=\prod_{i=0}^{a-1}(1-xq^i)$, and for $a \leq 0$
\begin{align}
\begin{split}
    \hat{A}(x_1,x_2,\hat{y}) &=  (-q^{1/2})^{-a} x_1 (x_2;q)_{1-a} +(q^{-a} x_2 - 1 + \hat{y}) \hat{y}^{-a}, \\
    \hat{A}(x_1=c_1x,x_2=c_2x,\hat{y}) &=  (-q^{1/2})^{-a} x c_1 (x c_2;q)_{1-a} +(q^{-a} x c_2 - 1 + \hat{y}) \hat{y}^{-a}, \\
    \hat{A}(x_1=x,x_2=x,\hat{y}) &=  (-q^{1/2})^{-a} x (x;q)_{1-a} +(q^{-a} x - 1 + \hat{y}) \hat{y}^{-a}, \\
    \hat{A}(x_1=x,x_2=-x,\hat{y}) &=  (-q^{1/2})^{-a} x (x;q)_{1-a} +(-q^{-a} x - 1 + \hat{y}) \hat{y}^{-a}.
\end{split}
\end{align}
Corresponding classical A-polynomials, for  $a \geq 1$, are
\begin{align}
\begin{split}
    A(x_1,x_2,\hat{y}) &= (-1)^a x_1 y^a + (1 - x_2)^{a - 1} y - (1 - x_2)^a,  \\
    A(x_1=c_1x,x_2=c_2x,y) &= (-1)^a x c_1 y^a + (1 - x c_2)^{a - 1} y - (1 - x c_2)^a, \\
    A(x_1=x,x_2=x,y) &=   (-1)^a x y^a + (1 - x)^{a - 1} y - (1 - x)^a, \\
    A(x_1=x,x_2=-x,y) &= (-1)^a x y^a + (1 + x;q)^{a - 1} y - (1 + x)^a, \\
\end{split}
\end{align}
and for $a \leq 0$
\begin{align}
\begin{split}
    A(x_1,x_2,y) &= (-1)^{-a} x_1 (1 - x_2)^{1-a} +(x_2 - 1 + y) y^{-a}, \\
    A(x_1=c_1x,x_2=c_2x,y) &=   (-1)^{-a} x c_1 (1 - x c_2)^{1-a} +(x c_2 - 1 + y) y^{-a}, \\
    A(x_1=x,x_2=x,y) &=  (-1)^{-a} x (1 - x)^{1-a} +(x - 1 + y) y^{-a}, \\
    A(x_1=x,x_2=-x,y) &= (-1)^{-a} x (1 + x)^{1-a} +(-x - 1 + y) y^{-a}.
\end{split}
\end{align}
The parametrizations of classical curves are however universal for all $a\in\mathbb{Z}$ and take form
\begin{align}
\begin{split}
    x(t) &= \frac{c_1^{a-2}(t+(-1)^ac_1)}{t^a}, \quad y(t) =  \frac{t^a-c_1^{a-2}c_2t+(-1)^{a+1}c_1^{a-1}c_2}{(-1)^{a+1}c_1t^{a-1}}, \\
    x(t) &= \frac{(t+(-1)^a)}{t^a}, \quad y(t) =  \frac{t^a-t+(-1)^{a+1}}{(-1)^{a+1}t^{a-1}}, \qquad \textrm{for}\ c_1 = c_2 = 1, \\
    x(t)& = \frac{(t+(-1)^a)}{t^a}, \quad y(t) =  \frac{t^a+t+(-1)^a}{(-1)^{a+1}t^{a-1}}, \qquad \textrm{for}\ c_1 = - c_2 = 1.    \label{param-a000}
\end{split}
\end{align}

%-

\subsubsection*{Family ${a\ 0 \brack 0\ 1}$}

Furthermore, we consider the family 
\begin{equation}
    C=\begin{bmatrix}
    a & 0 \\
    0 & 1
    \end{bmatrix} \label{family-a1}
\end{equation}
with $a\in \mathbb{Z}$. This family is related to (\ref{family-a0}) by the reciprocity operation (\ref{reciprocal}), and one can easily derive the form of quantum and classical A-polynomials from those given above. Nonetheless, for convenience, and in order to illustrate how the reciprocity operation works, we provide these A-polynomials explicitly. Quantum A-polynomials, for $a \geq 1$, take form
\begin{align}
\begin{split}
    \hat{A}(x_1,x_2,\hat{y}) = & (-q^{1/2})^a x_1 (q^{1/2} x_2;q)_a \hat{y}^a + (1 - q^{1/2} x_2) \hat{y} - 1,  \\
    \hat{A}(x_1=c_1x,x_2=c_2x,\hat{y}) = & (-q^{1/2})^a x c_1 (q^{1/2} x c_2;q)_a \hat{y}^a + (1 - q^{1/2} x c_2) \hat{y} - 1, \\
    \hat{A}(x_1=x,x_2=x,\hat{y}) = & (-q^{1/2})^a x (q^{1/2} x;q)_a \hat{y}^a + (1 - q^{1/2} x) \hat{y} - 1, \\
    \hat{A}(x_1=x,x_2=-x,\hat{y}) = & (-q^{1/2})^a x (-q^{1/2} x;q)_a \hat{y}^a + (1 + q^{1/2} x) \hat{y} - 1 ,
\end{split}
\end{align}
and for $a \leq 0$ take form
\begin{align}
\begin{split}
    \hat{A}(x_1,x_2,\hat{y}) = & (q^{1/2} x_2;q)_{a+1} \hat{y}^{a+1} + (-q^{-1/2})^a x_1 - (q^{1/2} x_2;q)_a \hat{y}^a,  \\
    \hat{A}(x_1=c_1x,x_2=c_2x,\hat{y}) = & (q^{1/2} x c_2;q)_{a+1} \hat{y}^{a+1} + (-q^{-1/2})^a x c_1 - (q^{1/2} x c_2;q)_a \hat{y}^a, \\
    \hat{A}(x_1=x,x_2=x,\hat{y}) = & (q^{1/2} x;q)_{a+1} \hat{y}^{a+1} + (-q^{-1/2})^a x - (q^{1/2} x;q)_a \hat{y}^a, \\
    \hat{A}(x_1=x,x_2=-x,\hat{y}) = & (-q^{1/2} x;q)_{a+1} \hat{y}^{a+1} + (-q^{-1/2})^a x - (-q^{1/2} x;q)_a \hat{y}^a.
\end{split}
\end{align}
In consequence, classical A-polynomials, for $a \geq 1$, read
\begin{align}
\begin{split}
    A(x_1,x_2,y) = & (-1)^a x_1 (1 - x_2)^a y^a + (1 - x_2) y - 1,  \\
    A(x_1=c_1x,x_2=c_2x,y) = & (-1)^a x c_1 (1 - x c_2)^a y^a + (1 - x c_2) y - 1, \\
    A(x_1=x,x_2=x,y) = & (-1)^a x (1 - x)^a y^a + (1 - x) y - 1, \\
    A(x_1=x,x_2=-x,y) = & (-1)^a x (1 + x)^a y^a + (1 + x) y - 1, \\
\end{split}
\end{align}
and for $a \leq 0$ take form
\begin{align}
\begin{split}
    A(x_1,x_2,y) = & (1 - x_2)^{a+1} y^{a+1} + (-1)^a x_1 - (1 - x_2)^a y^a, \\
    A(x_1=c_1x,x_2=c_2x,y) = & (1 - x c_2)^{a+1} y^{a+1} + (-1)^a x c_1 - (1 - x c_2)^a y^a, \\
    A(x_1=x,x_2=x,y) = & (1 - x)^{a+1} y^{a+1} + (-1)^a x - (1 - x)^a y^a, \\
    A(x_1=x,x_2=-x,y) = & (1 + x)^{a+1} y^{a+1} + (-1)^a x - (1 + x)^a y^a.
\end{split}
\end{align}
The parametrizations, for all $a \in \mathbb{Z}$, take form
\begin{align}
\begin{split}
    x(t) &= \frac{c_1^{a-2}(t+(-1)^ac_1)}{t^a}, \quad y = \frac{(-1)^{a+1}c_1^{-1}t^{a+1}}{t^a-c_1^{a-2}c_2t+(-1)^{a+1}c_1^{a-1}c_2}, \\
    x(t) &= \frac{(t+(-1)^a)}{t^a}, \quad y = \frac{(-1)^{a+1}t^{a+1}}{t^a-t+(-1)^{a+1}}, \qquad \textrm{for}\ c_1 = c_2 = 1, \\
    x(t) &= \frac{(t+(-1)^a)}{t^a}, \quad y = \frac{(-1)^{a+1}t^{a+1}}{t^a+t+(-1)^a}, \qquad \textrm{for}\ c_1 = - c_2 = 1.
\end{split}
\end{align}

\subsubsection*{Reciprocal quivers}

The next family that yields A-polynomials of genus 0 is characterized by
\begin{equation}
    C=\begin{bmatrix}
    a & 0 \\
    0 & 1-a
    \end{bmatrix}\label{family-a1-a}
\end{equation}
It is sufficient to consider $a > 0$; the case $a \leq 0$ is equivalent to interchanging $x_1 \leftrightarrow x_2$. For $a>0$ quantum A-polynomials read
\begin{align}
\begin{split}
    \hat{A}(x_1,x_2,\hat{y}) &= \prod_{n=0}^{a-1} \left(q^a x_1 \hat{y}^a - q^{(a-n)(a-1)+1/2} x_2\right) + \\
    & \quad + q^{a/2} \prod_{n=0}^{a-2} \left(q^{n+a} x_1 \hat{y}^{a-1} - q^{(a-n)(a-1)+1/2} x_2\right) (1-\hat{y}) (-\hat{y})^{a-1}, \\
    \hat{A}(x_1=c_1x,x_2=c_2x,\hat{y}) &= x \prod_{n=1}^{a} \left(q^{(2 n - 1)/2} c_1 \hat{y}^a - c_2\right)+\\
    & \quad + (1-\hat{y}) (-q^{-1/2} \hat{y})^{a-1} \prod_{n=1}^{a - 1} \left(q^{n - a + 1/2} c_1 \hat{y}^{a-1} - c_2\right),\\
    \hat{A}(x_1=x,x_2=x,\hat{y}) &= x \prod_{n=1}^{a} \left(q^{(2 n - 1)/2} \hat{y}^a - 1\right)+  \\
    & \quad+ (1-\hat{y}) (-q^{-1/2} \hat{y})^{a-1} \prod_{n=1}^{a - 1} \left(q^{n - a + 1/2} \hat{y}^{a-1} - 1\right),\\
    \hat{A}(x_1=x,x_2=-x,\hat{y}) &= x \prod_{n=1}^{a} \left(q^{(2 n - 1)/2} \hat{y}^a + 1\right)+ \\
    & \quad+ (1-\hat{y}) (-q^{-1/2} \hat{y})^{a-1} \prod_{n=1}^{a - 1} \left(q^{n - a + 1/2} \hat{y}^{a-1} + 1\right),
\end{split}
\end{align}
and corresponding classical A-polynomials
\begin{align}\label{ClasicalAPolYReciprocal}
\begin{split}
    A(x_1,x_2,y) &= (x_1 y^a - x_2)^a + (1 - y) (-y)^{a-1} (x_1 y^{a-1} - x_2)^{a-1}, \\
    A(x_1=c_1x,x_2=c_2x,y) &=  x (c_1 y^a - c_2)^a + (1 - y) (-y)^{a-1} (c_1 y^{a-1} - c_2)^{a-1}, \\
    A(x_1=x,x_2=x,y) &= x (y^a - 1)^a + (1 - y) (-y)^{a-1} (y^{a-1} - 1)^{a-1} =\\
       &= (y - 1)^a [x \delta_a(y)^a - (-y)^{a-1} \delta_{a-1}(y)^{a-1}],  \\
    A(x_1=x,x_2=-x,y) &= x (y^a + 1)^a + (1 - y) (-y)^{a-1} (y^{a-1} + 1)^{a-1},
\end{split}
\end{align}
 where
 \be
  \delta_a(y) = \sum_{i=0}^{a-1} y^i = \frac{1-y^a}{1-y}.    \label{delta-a}
 \ee
We find the following parametrizations 
\begin{align}
\begin{split}
    x(t) &= (t-1) (-t)^{a-1} \frac{(c_1 t^{a-1} - c_2)^{a-1}}{(c_1 t^a - c_2)^a}, \quad y(t) = t, \\
    x(t) &= (-t)^{a-1}\frac{\delta_{a-1}(t)^{a-1}}{\delta_a(t)^a}, \quad y(t) = t, \qquad\textrm{for}\  c_1 = c_2 = 1, \\
    x(t) &= (t-1) (-t)^{a-1} \frac{(t^{a-1} + 1)^{a-1}}{(t^a + 1)^a}, \quad y(t) = t,  \qquad\textrm{for}\ c_1 = - c_2 = 1.
\end{split}
\end{align}
Remarkably,  specializing $x_1=x_2$ the classical A-polynomial factorizes and the parametrization only works for one of the factors. We can obtain a parametrization for the second factor, which encode the asymptotics for the quiver generating function \eqref{ClasicalAPolYReciprocal}, using the transformation $t \rightarrow (c_2 t - 1)/(c_1 t - 1)$
\begin{align}
\begin{split}
    x(t) &=  t \frac{\left[(c_1 t - 1)(c_2 t - 1) (-\Delta_{a-1}(t))\right]^{a - 1}}{\Delta_a(t)^a}, \quad y(t) = \frac{c_2 t - 1}{c_1 t - 1}, \\
    x(t) &= -\frac{t [(a - 2) t + 1]^{a-1}}{[(1 - a) t - 1]^a}, \quad y(t) = 1, \qquad\textrm{for}\  c_1 = c_2 = 1, \\
    x(t) &= (t-1) (-t)^{a-1} \frac{(t^{a-1} + 1)^{a-1}}{(t^a + 1)^a}, \quad y(t) =\frac{1+ t}{1-t}, \qquad\textrm{for}\  c_1 = - c_2 = 1,
\end{split}
\end{align}
where
\be
    \Delta_a(t) =  \frac{c_1 (c_2 t - 1)^a - c_2 (c_1 t - 1)^a}{(-1)^{a + 1} (c_2 - c_1)}  
    = 1 - c_1 c_2 \sum_{n=2}^a \binom{a}{n} (-t)^n \sum_{m=0}^{n-2} c_1^m c_2^{n-2-m}.
\ee

\subsubsection*{Diagonal quivers}

The last family that yields A-polynomials of genus 0 is that of diagonal quivers
\begin{equation}
    C=\begin{bmatrix}
    a & 0 \\
    0 & a
    \end{bmatrix}  \label{family-aa}
\end{equation}
We focus on $a\geq 0$, and the results for $a < 0$ can be obtained from $a > 1$ by reciprocity operation (\ref{reciprocal}). For given $a$, let us provide first expressions for quantum and classical A-polynomials and their parametrization. While for all families discussed earlier quantum A-polynomials have rather simple form, for diagonal quivers quantum A-polynomial is more complicated and takes form
\begin{equation}
\begin{aligned}
\widehat{A}(x_1,x_2,\hat y) = &\ \widehat{y} - (1 - (-1)^ax_1 \widehat{P})^{-1} (1+(-1)^ax_1\widehat{Q})
(\widehat{P}'_0 - (1 - (-1)^ax_1 \widehat{P})^{-1} \times \\
&\ (1+(-1)^ax_1\widehat{Q})),  \label{aa_Ahat}
\end{aligned}
\end{equation}
where $\widehat{P}, \widehat{Q}$, and $\widehat{P}'_0$ are certain operators in $x_1, x_2$, and $\hat y$. 
Below we will explain the structure of these operators and provide more details about a derivation of  (\ref{aa_Ahat}). However, let us first briefly summarize other results. In the $q=1$ limit (\ref{aa_Ahat}) yields the following expression for the classical resultant for $a>1$
\begin{equation}
A(x_1,x_2,y) =  \left( P+Q \right)  \left( y-1 \right) +
x_{{1}}x_{{2}} \left( P-Q \right) {y}^{a} + (-1)^a \left( x_{{1}}+x_{{2}} \right) {y}^{a},   \label{Adiag-clas}
\end{equation}
where
\begin{align}
\begin{split}
P =&\ \sum_{k=0}^{\lfloor \frac{a-1}{2} \rfloor}{a-1-k\choose k} \left(-x_1x_2y^a+y+1 \right) ^{a-1-2\,k}  {(-y)}^{k}, \\
Q =&\ \sum_{k=0}^{\lfloor \frac{a-1}{2} \rfloor}{a-2-k\choose k}\left( -x_1x_2y^a+y+1 \right) ^{a-2-2\,k}  {(-y)}^{k+1}.  \label{PQ}
\end{split}
\end{align}
For example
\begin{align}
\begin{split}
A(x_1,x_2,y)|_{a=2} &= x_1^2x_2^2y^4 - x_1x_2y^3 + ((-2x_2-1)x_1 - x_2)y^2 - y + 1, \\
A(x_1,x_2,y)|_{a=3} &= x_1^3x_2^3y^9 - 2x_1^2x_2^2y^7 - 3x_1^2x_2^2y^6 + x_1x_2y^5 + x_1x_2y^4 + \\
&\qquad + \big((3x_2-1) x_1 - x_2\big)y^3 + y - 1.
\end{split}
\end{align}
Furthermore, if we identify $x_1=x_2=x$, the classical A-polynomials factorizes
\be
A(x,y) = A_1(x,y) A_2(x,y),  \label{Ax1x2-diagonal}
\ee
where (depending on the parity of $a$)
\be
A_1(x,y) = \left\{\begin{array}{cl}
(xy^n-1)^2 - y = 0 & \qquad \textrm{for}\ a=2n \\
(xy^n-1)^2 y - 1 = 0 & \qquad \textrm{for}\ a=2n+1 
 \end{array} \right.
\ee
while $A_2(x,y)$ for $a=2,3,4,5\ldots$ reads
\begin{equation}
\begin{aligned}
A_2(x,y)|_{a=2} &= xy+1, \\
A_2(x,y)|_{a=3} &= x^2y^3+xy^2-1, \\
A_2(x,y)|_{a=4} &= x^3y^6+x^2y^4-xy^3-xy^2-1, \\
A_2(x,y)|_{a=5} &= x^4y^{10}+x^3y^8-x^2y^6-2x^2y^5-xy^4-xy^3+1.
\end{aligned}
\end{equation}

The curves arising as $(x_1=c_1x, x_2=c_2x)$ specialization of (\ref{Adiag-clas}) are parametrized as follows
\begin{align}
\begin{split}
    x(t) &= \frac{(t-1) (-t)^{a-1} (c_1 t^{a-1} - c_2)^{a-1}}{(c_1 t^a - c_2)^a}, \quad y(t) = \frac{(c_1 t^a - c_2)^2}{t (c_1 t^{a - 1} - c_2)^2}, \\
    x(t) &= t (-t)^{a-1}\frac{\delta_{a-1}(t)^{a-1}}{\delta_a(t)^a}, \quad y(t) = \frac{\delta_a(t)^2}{t \delta_{a-1}(t)^2}, \qquad\textrm{for}\   c_1 = c_2 = 1, \\
    x(t) &= \frac{(t-1) (-t)^{a-1} (t^{a-1} + 1)^{a-1}}{(t^a + 1)^a}, \quad y(t) = \frac{(t^a + 1)^2}{t (t^{a - 1} + 1)^2}, \qquad\textrm{for}\   c_1 = - c_2 = 1,
\end{split}
\end{align}
where $\delta_a(t)$ is defined in (\ref{delta-a}). Similarly to the factorization of the classical A-polynomial in the reciprocal quiver case when $x_1=x_2$, the previous parametrization works for $A_2(x,y)$. We can find a parametrization for $A_1(x,y)$ by the same transformation $t\rightarrow (c_2 t-1)/(c_1 t-1)$
\begin{align}
\begin{split}
    x(t) &= t \frac{\left[(c_1 t - 1)(c_2 t - 1) (-\Delta_{a-1}(t))\right]^{a - 1}}{\Delta_a(t)^a}, y(t) = \frac{\Delta_a(t)^2}{(c_1 t - 1) (c_2 t - 1) \Delta_{a - 1}(t)^2}, \\
    x(t) &= -\frac{t [(a - 2) t + 1]^{a-1}}{[(1 - a) t - 1]^a}, \quad y(t) = \left[\frac{(a-1) t +1}{(a-2) t +1} \right]^2, \qquad\textrm{for}\   c_1 = c_2 = 1 \\
    x(t) &= -t \frac{\left[(t - 1)(t + 1) (-\tilde{\Delta}_{a-1}(t))\right]^{a - 1}}{\tilde{\Delta}_a(t)^a}, \quad y(t) = - \frac{\tilde{\Delta}_a(t)^2}{(t - 1) (t + 1) \tilde{\Delta}_{a - 1}(t)^2}, \\ 
    & \qquad\textrm{for}\   c_1 = - c_2 = 1,
\end{split}
\end{align}
where
\begin{equation}
    \tilde{\Delta}_a(t) = \frac{(t + 1)^a + (1 - t)^a}{2}.
\end{equation}

Let us derive now the quantum A-polynomial (\ref{aa_Ahat}). For a quiver (\ref{family-aa}) quantum Nahm equations take form
\begin{equation}\label{QNahm_diag}
\widehat{z}_1P_C =   (1-(-1)^ax_1\widehat{z}_1^a)P_C,\qquad
\widehat{z}_2P_C =   (1-(-1)^ax_2\widehat{z}_2^a)P_C,
\end{equation}
and as usual $\widehat{y} = \widehat{z}_1\widehat{z}_2$. Rewriting the second equation $\widehat{z}_2 P_C = (1 - (-1)^a x_2 (\widehat{z}_1^{-1}\widehat{y})^a )P_C$, and then $\widehat{z}_1^{a}\widehat{y} P_C = (\widehat{z}_1^{a+1} - (-1)^a x_2 \widehat{z}_1\widehat{y}^a)P_C$, we get
\begin{equation}
\widehat{z}_1(\widehat{z}_1^a P_C) = 
(\widehat{y}\widehat{z}_1^a + (-1)^ax_2 \widehat{z}_1\widehat{y}^a) P_C.
\end{equation}
Substituting $\widehat{z}_1^a P_C$ from the first equation in (\ref{QNahm_diag}) we get
\begin{equation}\label{z1eq}
\widehat{z}_1^2 P_C = ((-qx_1x_2\widehat{y}^a+\widehat{y}+1) \widehat{z}_1 - \widehat{y}) P_C,
\end{equation}
and an analogous equation for $\widehat{z}_2$. We can multiply this equation from the left by $\widehat{z}_1$ as many times as we want, and at each step we can resolve $\widehat{z}_1^k$ in terms of a linear combination of $\widehat{z}_1$ with coefficients in $x_1,x_2,\widehat{y}$. 
\begin{figure}[h]
    \centering
    \includegraphics[scale=0.5]{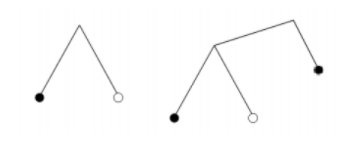}
    \caption{The graphs $G_2$ for $a = 2$ (left) and $G_3$ for $a = 3$ (right).}
    \label{initialgraphs}
\end{figure}
Iterating, we obtain the expression
\begin{equation}\label{reducer}
\widehat{z}_1^a P_C = (\widehat{P}\widehat{z}_1 + \widehat{Q}) P_C
\end{equation}
where $\widehat{P}$ and $\widehat{Q}$ are operators, whose form (for a given $a$) we encode in a tree-like two-valent graph $G_a$, as follows. A graph $G_a$ has a root (drawn on top), and an interchanging pattern of black and white dots at the bottom. The graphs for $a=2$ and $a=3$, are shown in fig. \ref{initialgraphs}, and graphs $G_a$  for $a>3$ are constructed recursively: a root of $G_a$ is connected to $G_{a-1}$ on the left and to $G_{a-2}$ on the right, see fig. \ref{graphs4and5} for examples for $a=4$ and $a=5$. Clearly, the numbers of black and white circles are given by Fibonacci numbers. Furthermore, we define
\begin{equation}
\widehat{P}_0 = -qx_1x_2\widehat{y}^a+\widehat{y}+1,\qquad \widehat{Q}_0 = -\widehat{y}.
\end{equation}
Now, in a graph $G_a$, from a root one can travel to each dot along a ``path'' that consists of right and left turns (we take a turn at each two-valent vertex). We associate to such a path an operator that we call ``${\rm word} =  {\rm word}(\widehat{P}_0,\widehat{Q}_0)$'', by assigning (in the order of traveling from the root to a given dot) the operator $\widehat{P}_0$ to each left turn, and $\widehat{Q}_0$ to each right turn. Then
\begin{equation}\label{PhatQhat}
\begin{aligned}
\widehat{P} = &\ \sum_{\{\text{all paths in $G_a$ ending with }\bullet\} } {\rm word}(\widehat{P}_0,\widehat{Q}_0), \\
\widehat{Q} = &\ \sum_{\{\text{all paths in $G_a$ ending with }\circ\} } {\rm word}(\widehat{P}_0,\widehat{Q}_0).
\end{aligned}
\end{equation}
For example, for $a=2$ we obtain $\widehat{P} = \widehat{P}_0$ and $\widehat{Q} = \widehat{Q}_0$. For $a = 3$ we get $\widehat{P} = \widehat{P}_0\widehat{P}_0 + \widehat{Q}_0$ and $\widehat{Q} = \widehat{P}_0\widehat{Q}_0$.
\begin{figure}[h]
    \centering
    \includegraphics[scale=0.5]{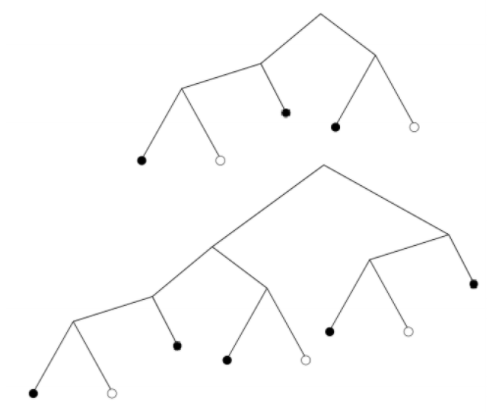}
    \caption{Graphs $G_4$ for $a = 4$ (obtained by gluing $G_3$ and $G_2$ to a new root, top) and $G_5$ for $a = 5$ (obtained by gluing $G_4$ and $G_3$ to a new root, bottom)}
    \label{graphs4and5}
\end{figure}
Having constructed $\widehat{P}$ and $\widehat{Q}$, we plug (\ref{reducer}) into the first Nahm equation (\ref{QNahm_diag}) for $\widehat{z}_1 P$, which yields
\begin{equation}
\widehat{z}_1 P_C = (1 - (-1)^a x_1 (\widehat{P} \widehat{z}_1 + \widehat{Q})) P_C,  
\end{equation}
so that we get
\begin{equation}\label{forz1}
%((1+(-1)^ax_1 \widehat{P})\widehat{z}_1)P_C = &\ (1 - (-1)^ax_1 \widehat{Q})P_C, \\
\widehat{z}_1 P_C = (1+(-1)^ax_1 \widehat{P})^{-1} (1 - (-1)^ax_1 \widehat{Q})P_C.
\end{equation}
Recall now that $\widehat{z}_1P_C$ and $\widehat{z_2} P_C$ are related by linear equation, and if we multiply (\ref{z1eq}) by $\widehat{z}_2$ from the left, and plug $\widehat{y}$ where possible, we get
\begin{equation}\label{forz2}
\begin{aligned}
\widehat{y}\widehat{z}_1 P_C = &\
((q^2x_1x_2\widehat{y}^a-\widehat{y}-1) \widehat{y} - \widehat{y}z_2) P_C, \\
\widehat{z}_2 P_C = &\ (\widehat{P}'_0 - \widehat{z}_1) P_C,
\end{aligned}
\end{equation}
where $\widehat{P}'_0 = -q^2 x_1 x_2 y^a + y + 1$. Finally, combining the last lines in (\ref{forz1}) and (\ref{forz2}) into $\widehat{y}-\widehat{z}_1\widehat{z}_2$, we can formally write the $\widehat{A}$ polynomial as follows
%an element in the Ore algebra
\begin{equation}
\begin{aligned}
\widehat{A}(x_1,x_2,\hat y) = &\ \widehat{y} - (1 - (-1)^ax_1 \widehat{P})^{-1} (1+(-1)^ax_1\widehat{Q})
(\widehat{P}'_0 - (1 - (-1)^ax_1 \widehat{P})^{-1} \times \\
&\ (1+(-1)^ax_1\widehat{Q})),
\end{aligned}
\end{equation}
as we announced in (\ref{aa_Ahat}). In the $q=1$ limit this reduces to the A-polynomial (\ref{Adiag-clas}).

\subsubsection*{Diagonal case ${2\ 0 \brack 0\ 2}$}

As a special case of the above family of diagonal quivers -- which we will also analyze via topological recursion in what follows -- let us consider $a=2$, so that 
\begin{equation}
    C=\begin{bmatrix}
    2 & 0 \\
    0 & 2
    \end{bmatrix}  \label{diagonal-22}
\end{equation}
In this case quantum A-polynomials take form
\begin{align}
\begin{split}
    \hat{A}(x_1,x_2,\hat{y}) &= (q^3 x_1 x_2 \hat{y}^2 - 1) (q^2 x_1 x_2 \hat{y}^2 - 1) - (q x_2 \hat{y} + 1) (q x_1 \hat{y} + 1) \hat{y}, \\
    \hat{A}(x_1=c_1x,x_2=c_2x,\hat{y}) &= (q^3 x^2 c_1 c_2 \hat{y}^2 - 1) (q^2 x c_1 c_2 \hat{y}^2 - 1) - (q x c_2 \hat{y} + 1) (q x c_1 \hat{y} + 1) \hat{y}, \\
    \hat{A}(x_1=x,x_2=x,\hat{y}) &= (q^3 x^2  \hat{y}^2 - 1) (q^2 x  \hat{y}^2 - 1) - (q x  \hat{y} + 1) (q x  \hat{y} + 1) \hat{y} =\\
    & = [q x \hat{y} + 1]  [(q x \hat{y} - 1) (q^2 x^2 \hat{y}^2 - 1) - (q x \hat{y} + 1) \hat{y}], \\
    \hat{A}(x_1x,x_2=-x,\hat{y}) &= (q^3 x^2 \hat{y}^2 + 1) (q^2 x \hat{y}^2 + 1) - (q x \hat{y} - 1) (q x \hat{y} - 1) \hat{y}.
\end{split}
\end{align}
Note that for $x_1=x_2=x$ the above quantum curve factorizes. Furthermore, classical A-polynomials read
\begin{align}
\begin{split}
    A(x_1,x_2,y) &= (x_1 x_2 y^2 - 1)^2 - (x_2 y + 1) ( x_1 y + 1) y, \\
    A(x_1=c_1x,x_2=c_2x,y) &=  (x^2 c_1 c_2 y^2 - 1)^2 - (x c_2 y + 1) (x c_1 y + 1) y, \\
    A(x_1=x,x_2=x,y) &= (x y + 1)^2 [(x y - 1)^2 - y], \\
    A(x_1=x,x_2=-x,y) &=  (x^2 y^2 + 1)^2 + (x^2 y^2 - 1) y. \\
\end{split}
\end{align}
One parametrization of these curves takes form
\begin{align}
\begin{split}
    x(t) = & (t-1) (-t) \frac{(c_1 t - c_2)}{(c_1 t^2 - c_2)^2}, \quad y(t) = \frac{(c_1 t^2 - c_2)^2}{t (c_1 t - c_2)^2}, \\
    x(t) = & \frac{-t}{(t + 1)^2}, \quad y(t) = \frac{(t + 1)^2}{t},\qquad\textrm{for}\   c_1 = c_2 = 1 \\
    x(t) = & (t-1) (-t) \frac{(t + 1)}{(t^2 + 1)^2}, \quad y(t) = \frac{(t^2 + 1)^2}{t (t + 1)^2}, \qquad\textrm{for}\   c_1 = -c_2 = 1,
\end{split}
\end{align}
and another one, obtained from the above one after the transformation $t \mapsto (c_2 t - 1)/(c_1 t - 1)$
\begin{align}
\begin{split}
    x(t) = & -t \frac{\left(c_1 t-1\right) \left(c_2 t-1\right)}{\left(c_1 c_2 t^2-1\right)^2}, \quad y(t) = \frac{\left(c_1 c_2 t^2-1\right)^2}{\left(c_1 t-1\right) \left(c_2 t-1\right)}, \\
    x(t) = & -\frac{t}{(t+1)^2}, \quad y(t) = (t+1)^2, \qquad\textrm{for}\   c_1 = c_2 = 1, \\
    x(t) = & (t-1) (-t) \frac{(t + 1)}{(t^2 + 1)^2}, \quad y(t) = \frac{(t^2 + 1)^2}{t (t + 1)^2}, \qquad\textrm{for}\   c_1 = -c_2 = 1.
\end{split}
\end{align}

%***

\subsubsection*{Other 2-vertex quivers}

Above we identified all 2-vertex quivers whose A-polynomials have genus 0. Now we are going to determine A-polynomials of genus $g>0$. We find that there is a finite number of matrices $C$ which yield such A-polynomials. To present such matrices, we recall that apart from a change in farming (\ref{framing}), also the reciprocity operation (\ref{reciprocal}) does not change the genus. Moreover, for diagonal matrices (\ref{diagonal-ab}), we find that also refined versions of the reciprocal operation, which just sends $a\mapsto 1-a$ or $b\mapsto 1-b$, leaves the genus of the A-polynomial invariant. For that reason, it is sufficient to identify matrices $C$ with positive $a$ and $b$, which yield A-polynomials of a genus $g>0$. Therefore, having screened a wide ensemble of positive $a$ and $b$, we find that the resulting A-polynomial has genus $g>0$ for pairs $(a,b)$ such that $a=2,\ldots,g+1$, and for a given $a$ the value of $b$ is in the following range
\begin{equation}
\text{genus $= g$, $(a,b):$}\quad 
\begin{tabular}{c}
$(2,2g+1), (2,2g+2)$ \\
$(3,3g-2), (3,3g-1), (3,3g)$ \\
$(4,4g-7), (4,4g-6), (4,4g-5), (4,4g-4)$ \\
$\vdots$ \\
$\underbrace{(g+1,g+2),\dots, (g+1,2(g+1))}_{g+1}$
\end{tabular}
\end{equation}
In particular, the values of $(a,b)$ that yield a curve of a given genus for small values of $g$ are as follows
\begin{equation}
g = 1:
\begin{tabular}{c}
$(2,3), (2,4)$,
\end{tabular}
g = 2:
\begin{tabular}{c}
$(2,5), (2,6)$ \\
$(3,4), (3,5), (3,6)$, \\
\end{tabular}
g = 3:
\begin{tabular}{c}
$(2,7), (2,8)$ \\
$(3,7), (3,8), (3,9)$ \\
$(4,5), (4,6), (4,7), (4,8)$
\end{tabular}
\end{equation}

There is no general formula for the form of the algebraic curve $A(x,y)=0$, or its parametrization, even for the diagonal case (\ref{diagonal-ab}) we are considering here. However, we can provide explicitly the form of $A(x,y)$ for some low values of $a$ and $b$. In particular, taking into account the transformations mentioned above that preserve the genus, and understanding that $x_1=c_1x$ and $x_2=c_2x$, in table \ref{tab-g1} we list all curves of genus 1 together with values of $(a,b)$ for which they arise.
\begin{table}
\begin{equation}
\begin{array}{c|c}
(a,b) & A(x_1,x_2,y) \\
\hline
(3,2) &\ {y}^{6}x_1^2x_2^3-2\,{y}^{4}x_{{1}}{x_{{2}}}^{2}-{y}^{
4}x_{{1}}x_{{2}}-3\,{y}^{3}x_{{1}}x_{{2}}-{y}^{3}x_{{1}}+{y}^{2}x_{{2}
}+y-1 \\
(4,2)  &\ {y}^{8}x_1^2x_2^4-3\,{y}^{5}x_{{1}}{x_{{2}}}^{2}-{y}^{
5}x_{{1}}x_{{2}}-2\,{y}^{4}x_{{1}}{x_{{2}}}^{2}-4\,{y}^{4}x_{{1}}x_{{2
}}-{y}^{4}x_{{1}}-{y}^{2}x_{{2}}-y+1 \\
(-1,-2)  &\ {y}^{6}-{y}^{5}-{y}^{4}x_{{1}}+3\,{y}^{3}x_{{1}}x_{{2}}+2\,{y}^{2}x_1^2x_{{2}}-x_1^3x_2^2+{y}^{3}x_{{2}}+{y}^{2}x_{{1}}x_{{2}} \\
(-1,-3)  &\ {y}^{8}-{y}^{7}-{y}^{6}x_{{1}}-2\,{y}^{4}x_1^2x_{{2}}-4\,{y}^{
4}x_{{1}}x_{{2}}-3\,{y}^{3}x_1^2x_{{2}}+x_1^4x_2^2-{y}^{4}x_{{2}}-{y}^{3}x_{{1}}x_{{2}} \\
(3,-1)  &\ {y}^{6}x_1^2-2\,{y}^{4}x_{{1}}x_{{2}}-{y}^{4}x_{{1}}+3\,{y}^{3
}x_{{1}}x_{{2}}+{y}^{3}x_{{1}}+{y}^{2}{x_{{2}}}^{2}-y{x_{{2}}}^{2}-{x_
{{2}}}^{3} \\
(4,-1)  &\ {y}^{8}x_1^2+3\,{y}^{5}x_{{1}}x_{{2}}-2\,{y}^{4}x_{{1}}{x_{{2}
}}^{2}+{y}^{5}x_{{1}}-4\,{y}^{4}x_{{1}}x_{{2}}-{y}^{4}x_{{1}}-{y}^{2}x_2^3+yx_2^3+x_2^4 \\
(2,-2)  &\ {y}^{6}x_1^3+{y}^{5}x_1^2-{y}^{4}x_1^2-3\,{y}^
{3}x_{{1}}x_{{2}}-{y}^{3}x_{{2}}+2\,{y}^{2}x_{{1}}x_{{2}}+{y}^{2}x_{{2}}-{x_{{2}}}^{2} \\
(2,-3)  &\ {y}^{8}x_1^4+{y}^{7}x_1^3-{y}^{6}x_1^3-2\,{y}^{4}x_1^2x_{{2}}-4\,{y}^{4}x_{{1}}x_{{2}}-{y}^{4}x_{{2}}+3\,{y}^{3}x_{{1}}x_{{2}}+{y}^{3}x_{{2}}+{x_{{2}}}^{2}
\end{array} \nonumber
\end{equation}
\caption{All A-polynomials of genus $g=1$, $A(x,y) \equiv A(x_1,x_2,y)$, where $x_1=c_1x$ and $x_2=c_2x$, arising for diagonal matrices $C$ of size 2 with diagonal elements $a$ and $b$.}  \label{tab-g1}
\end{table}

Finally, in table \ref{tab-g-all} we present A-polynomials $A(x,y)$ for a number of matrices $C$ with lowest positive values of coefficients $a$ and $b$ (other than 0 and 1, as discussed earlier), and determine their genus, with identification of generating parameters $x_1=-x_2=x$. 

\begin{table}
\be
\begin{array}{c|c|c}
(a,b) & A(x,y) & g \\
\hline \hline
(2,2) &\ x^4y^4+x^2y^3+2x^2y^2-y+1 &\  0 \\
(2,3) &\ x^5y^6+2x^3y^4+x^2y^4+3x^2y^3+xy^3+xy^2+y-1 &\ 1 \\
(2,4) &\ x^6y^8+3x^3y^5+2x^3y^4+x^2y^5+4x^2y^4+xy^4-xy^2-y+1 &\ 1 \\
(2,5) &\ x^7y^{10}+2x^4y^6+4x^3y^6+5x^3y^5+x^2y^6+5x^2y^5+xy^5+xy^2+y-1 &\ 2 \\
(2,6) &\ x^8y^{12}+5x^4y^7+2x^4y^6+5x^3y^7+9x^3y^6+x^2y^7+6x^2y^6+xy^6-xy^2-y+1 &\ 2 \\
%(2,7), &\ x^9y^14+2x^5y^8+9x^4y^8+7x^4y^7+6x^3y^8+14x^3y^7+x^2y^8+ \\
%7x^2y^7+xy^7+xy^2+y-1, &\ g = 3
\hline
(3,3) &\ x^6y^9+2x^4y^7+3x^4y^6+x^2y^5+x^2y^4+3x^2y^3-y+1 &\  1 \\
(3,4) &\ -x^7y^{12}-3x^5y^9-2x^4y^8-3x^3y^6+x^2y^6+x^2y^5+4x^2y^4-xy^4-xy^3+y-1 &\  2 \\
(3,5) &\ -x^8y^{15}-2x^5y^{11}-5x^5y^{10}+4x^3y^7+3x^3y^6-x^2y^7 +\\
& -x^2y^6-5x^2y^5+xy^5-xy^3+y-1 &\  2 \\
(3,6) &\ -x^9y^{18}-5x^6y^{13}-3x^6y^{12}-2x^5y^{12}-5x^3y^8-4x^3y^7+
x^2y^8-3x^3y^6+ \\
& +x^2y^7+6x^2y^6-xy^6-xy^3+y-1
&\  2 \\
\hline
(4,4) &\  x^8y^{16}+3x^6y^{13}+4x^6y^{12}+3x^4y^{10}+5x^4y^9+6x^4y^8+x^2y^7+x^2y^6+ \\
& +x^2y^5+4x^2y^4-y+1 &\  0 \\
(4,5) &\ x^9y^{20}+4x^7y^{16}+6x^5y^{12}-3x^4y^{11}-5x^4y^{10}+4x^3y^8+x^2y^8+ \\
& +x^2y^7+x^2y^6+5x^2y^5+xy^5+xy^4+y-1 &\  3 \\
(4,6) &\ x^{10}y^{24}+3x^7y^{18}-5x^5y^{14}-11x^5y^{13}-2x^5y^{12}+3x^4y^{12}+5x^3y^9+ \\
& +4x^3y^8+x^2y^9+x^2y^8+x^2y^7+6x^2y^6+xy^6-xy^4-y+1 &\  3 \\
\hline
(5,5) &\ x^{10}y^{25}+4x^8y^{21}+5x^8y^{20}+6x^6y^{17}+11x^6y^{16}+10x^6y^{15}+4x^4y^{13}+7x^4y^{12}+ \\
& +9x^4y^{11}+10x^4y^{10}+x^2y^9+x^2y^8+x^2y^7+x^2y^6+5x^2y^5-y+1 &\  0 \\
(5,6) &\ -x^{11}y^{30}-5x^9y^{25}-10x^7y^{20}+2x^6y^{18}-10x^5y^{15}-4x^4y^{14}-7x^4y^{13}-9x^4y^{12} \\
& -5x^3y^{10}+x^2y^{10}+x^2y^9+x^2y^8+x^2y^7+6x^2y^6-xy^6-xy^5+y-1 &\  4 \\
\hline
(6,6) &\ x^{12}y^{36}+5x^{10}y^{31}+6x^{10}y^{30}+10x^8y^{26}+19x^8y^{25}+15x^8y^{24}+10x^6y^{21}+ \\
& +21x^6y^{20}+26x^6y^{19}+20x^6y^{18}+5x^4y^{16}+9x^4y^{15}+12x^4y^{14}+14x^4y^{13} + &\  0  \\
& +15x^4y^{12}+x^2y^{11}+x^2y^{10}+x^2y^9+x^2y^8+x^2y^7+6x^2y^6-y+1 
\end{array} \nonumber
\ee
\caption{A-polynomials $A(x,y)$, and their genus, for a number of diagonal matrices $C$ of size 2, with low positive values of coefficients $a$ and $b$ (other than 0 and 1, as discussed in the text), with identification of generating parameters $x_1=x_2=x$}  \label{tab-g-all}
\end{table}

%***************************************************************************************************

\subsection{Uniform quivers}    \label{ssec-uniform}

It is more difficult to classify A-polynomials for quivers with more than 2 vertices. In this work we do not provide such a general classification -- instead, we consider two families of quivers with arbitrary number of vertices $m$. The first family we consider is characterized by a constant $m\times m$ matrix whose each element is equal to $f$
\begin{equation}
    C =\left[ \begin{array} {cccc}
f & f & \cdots & f \\
f & f & \cdots & f \\
\vdots & \vdots &\ddots & \vdots\\
f & f & \cdots & f \\
\end{array} \right]  \label{C-uniform}
\end{equation}
These examples can be regarded as $C=0$ cases in framing $f$ (\ref{framing}). We refer to such quivers as uniform quivers. In what follows, we will illustrate that the topological recursion yields correct results for this family of quivers too.

For quiver matrices (\ref{C-uniform}) the generating function (\ref{QGF}) takes form
\be
P_C(x) = \sum_{(d_1, \dots, d_m) \geq 0} \frac{(-q^{1/2})^{f(d_1+ \dots + d_m)^2}}{(q;q)_{d_1} \cdots (q;q)_{d_m}} x_1^{d_1}\cdots x_m^{d_m}.
\ee
In this case the equation (\ref{QNE}) takes form
\be
\hat{z}_k P_C(x_1,\dots,x_m) = \left(1-(-q^{1/2})^f x_k \hat{y}^f\right) P_C(x_1,\dots,x_m),
\ee
so that, in view of (\ref{y-hat-z-hat}), the operator that annihilates the quiver generating series reads
%\be
%\hat{y} P_C(x_1,\dots,x_m) = \prod_{k=1}^m \left(1-(-q^{1/2})^c x_k \hat{y}^c\right) P_C(x_1,\dots,x_m)
%\ee
\be
     \hat{A}(x_1,\dots,x_m,\hat{y}) = \prod_{k=1}^m \left(1-(-q^{1/2})^f x_k \hat{y}^f\right) - \hat{y}. \label{QAPUQ}
\ee
Its classical limit, which agrees with the analysis of asymptotic equations (\ref{S0-quiver}), and equivalently with the resultant (\ref{resultant}), takes form
\be
     A(x_1,\dots,x_m,y) = \prod_{k=1}^m \left(1 - x_k (-y)^f\right) - y. \label{CAPUQ}
\ee
If we then set $x_i=x$ in the above expressions, we obtain the following quantum quiver A-polynomial
\be
\hat{A}(\hat{x}, \hat{y}) = (1 - (-1)^f q^{f/2} \hat{x} \hat{y}^f)^m - \hat{y},   \label{Ahat-xy-uniform}
\ee
and the classical A-polynomial curve
\begin{equation}
    A(x,y)=y-(1-(-1)^f x y^f)^m = 0.    \label{Axy-uniform}
\end{equation}
This curve has genus zero and its rational parametrization takes form 
\be
x(t)=\frac{1-t}{(-t^m)^f}, \qquad y(t)=t^m.    \label{uniform-param}
\ee

%***************************************************************************************************

\subsection{Yet another family of quivers}  \label{ssec-family}

We identify yet another, quite large family of quivers, of arbitrary size, whose A-polynomials are of genus 0, and whose structure is encoded in the following $m\times m$ matrix
\begin{equation}\label{C_yet_simple}
C = \left[\begin{array} {ccccc}
c+k_1+\delta_1 & c+k_1 & \cdots & \cdots & c+k_1 \\
c+k_1 & c+k_2+\delta_2 & c+k_2 & \cdots & c+k_2 \\
\vdots & c+k_2 &\ddots & \vdots & \vdots\\
\vdots & \vdots & \cdots & c+k_{m-1}+\delta_{m-1} & c+k_{m-1} \\
c+k_1 & c+k_2 & \cdots & c+k_{m-1} & 0 \\
\end{array} \right]
\end{equation}
with $c \geq 0,\ \delta_j\in \{0,1\}$, and a grading condition $\ k_1\geq \dots \geq k_{m-1} \geq 0$, such that $\sum_{i=1}^{m-1} \delta_i < \sum_{i=1}^{m-1} k_i$. For such quivers we can write down classical A-polynomials equations explicitly, from the form of Nahm equations. Namely, each row of (\ref{C_yet_simple}) corresponds to a particular $z_i$, and we plug $y=\prod_{i=1}^m z_i$ to the left hand side of each equation. Therefore, we get $z_i$ as a function of $(x_i,y)$, then move to the next row and repeat the procedure. Ultimately we find
\begin{equation}\label{A_cl}
A(x_1,\dots,x_m,y) =  y-\left(1-\prod_{i=1}^{m-1}Q_i^{c+k_i}x_{i+1}\right)\prod_{i=1}^{m-1}Q_i,
\end{equation}
where
\be
\left\{
\begin{matrix*}[l]
Q_1=\big(1-(-1)^{c+k_1+\delta_1}x_1y^{c+k_1}\big)^{(-1)^{k_1+\delta_1}}   \\
Q_j =\big(1-(-1)^{c+k_j+\delta_j}x_jy^{c+k_j}\prod_{s< j}Q_s^{k_s-k_j}\big)^{(-1)^{k_j+\delta_j}}\quad \textrm{for}\  j=2,\ldots ,(m-1)
\end{matrix*}
\right.
\ee
Upon specialization $x_i=c_i x$, the resulting A-polynomials are irreducible and have genus 0.

Let us consider a few special cases that belong to the family (\ref{C_yet_simple}). For $m=2$ we find two  families of quivers parametrized by $c\geq 0$. The first family takes form
\be
C = \begin{bmatrix} 
c & c \\
c & 0   
\end{bmatrix},     \label{family-ccc0}
\ee
and the corresponding resultant reads
\begin{equation}
A(x_1,x_2,y)  = y- \left( 1- \left( -1 \right) ^{c}x_{{1}}{y}^{c} \right)  \left( 1-x_{{2}} \left( 1- \left( -1 \right) ^{c}x_{{1}}{y}^{c} \right) ^{c} 
\right) .
\end{equation}
The second family is of the form
\be
C = \begin{bmatrix} 
c+1 & c \\
c & 0
\end{bmatrix} \label{family-c+1cc0}
\ee
and the corresponding resultant is
\begin{equation}
A(x_1,x_2,y) = \left( 1+ \left( -1 \right) ^{c}x_{{1}}{y}^{c} \right) ^{c+1}y-
\left( 1+ \left( -1 \right) ^{c}x_{{1}}{y}^{c} \right) ^{c}+x_{{2}}.
\end{equation}
Note that the data associated to such a quiver with $c=1$ (and in some specific framing) captures extremal invariants of the trefoil knot and certain Duchon paths \cite{Panfil:2018sis}.

Another example is the quiver of arbitrary size, whose matrix has a single non-zero entry in the top-left corner, $C_{1,1}=c>0$
\begin{equation}
C = \begin{bmatrix} 
c & 0 & 0 & \cdots \\
0 & 0 & 0 &\cdots \\
\vdots & & & \ddots
\end{bmatrix}
\end{equation}
In this case the resultant reads
\begin{equation}
A(x_1,\ldots,x_m,y) = \left( -1 \right) ^{c}x_{{1}}{y}^{c}+ \left( \prod_{i={2}}^{m-1}  \left( x_{i
}-1\right)  \right) ^{c-1} \left( y+\sum _{i=0}^{m-1}
\left( -1 \right) ^{i+1}e_{{i}} \left(x_1,\dots,x_m \right)  \right), 
\end{equation}
where $e_i(x_1,\dots,x_m)$ are elementary symmetric polynomials.

\section{Topological recursion and quantum curves}  \label{sec-tr-intro}

In this section we briefly review the formalism of topological recursion, as well as the construction of quantum curves that relies on the topological recursion. One can think of the topological recursion as a tool that assigns an infinite family of multi-differentials $\omega_{g,n}$ and free energies $F_g$ to a spectral curve. A spectral curve is a Riemann surface, together with a pair of functions $u=u(p)$ and $v=v(p)$ that parametrize it, i.e. they satisfy an equation
\be
A(u,v) = 0.   \label{Axy=0}
\ee
In the original formulation, $A(u,v)$ is a polynomial in $u$ and $v$ which both are in $\mathbb{C}$, and this equation defines an algebraic curve. In this work we will be rather interested in the case when the above is a polynomial equation in $\mathbb{C}^*$-valued variables, $x=e^u$ and $y=e^v$; this is analogous to the case of topological string theory or knot theory, where the curves (\ref{Axy=0}) are identified respectively with mirror curves or A-polynomials.   

There are a few other ingredients necessary to define the topological recursion. First, one needs to consider ramification points $p_i^*$, defined as zeros $du(p)|_{p=p_i^*}=0$, as well as the poles of $u$ of order at least 2. Locally, for $p$ in a neighborhood of a ramification point $p_i^*$, we consider a conjugate point $\overline{p}$, such that $u(\overline{p})=u(p)$. Second, we introduce a meromorphic differential $\omega_{0,1}=vdu$ and the Bergman kernel $\omega_{0,2}$, which is symmetric and has a double pole on a diagonal. For curves of genus zero, which will be of main interest in this work, the Bergman kernel takes form
\be
\omega_{0,2} = \frac{dp dq}{(p-q)^2}.  \label{Bergman}
\ee
Finally, we define the recursion kernel
\begin{equation}
K(t_1,t_2,t_3)=\frac{dE_{t_2, t_3}(t_1)}{\omega(t_2, t_3)},
\end{equation} 
with a one-form $dE_{t_2, t_3}(t_1)$ and a one-form $\omega(t_1, t_2)$ called ``the vertex'',
\begin{equation}
dE_{t_2, t_3}(t_1) = \frac{1}{2}\int_{\xi=t_3}^{t_2} \omega_{0,2}(t_1,\xi), \qquad \omega(t_1, t_2) = \omega_{0,1}(t_1)-\omega_{0,1}(t_2).
\end{equation}

With the above definitions, for $2g-2+n\geq 0$, the topological recursion assigns to a given spectral curve with simple ramification points, an infinite tower of multi-differentials defined as
\begin{equation}
\begin{aligned}
\omega_{g,n}(t_1,\dots,t_n)=&\sum_{i}\underset{t \rightarrow p^*_i}{\mathrm{Res}}  K(t_1,t,\overline{t})\Big( \omega_{g-1,n+1}(t,\overline{t},t_2,\dots,t_n)+ \\&
+ \sum'_{\substack{g_1+g_2=g \\ I_1 \cup I_2 = \{t_2,\dots,t_n\}}}\omega_{g_1,1+|I_1|}(t,I_1)\omega_{g_2,1+|I_2|}(\overline{t},I_2) \Big),
\end{aligned}
\end{equation}
where $\sum'$ excludes $\omega_{0,1}$and $\omega_{0,2}$ from the summation. For example
\begin{align}
\begin{split}
	\omega_{1,1}(t_1) &= \sum_i \underset{t \rightarrow p^*_i}{\mathrm{Res}} K(t_1,t,\overline{t}) \omega_{0,2}(t, \bar{t}), \\
	\omega_{0,3}(t_1, t_2, t_3) &= \sum_i \underset{t \rightarrow p^*_i}{\mathrm{Res}} K(t_1,t,\overline{t})\big(\omega_{0,2}(t, t_2) \omega_{0,2}(\bar{t}, t_3) + \omega_{0,2}(\bar{t}, t_2) \omega_{0,2}(t, t_3)\big).
\end{split}   \label{w11-w03}
\end{align}
$\omega_{g,n}$ are symmetric differentials of $n$'th order that share many nice properties; in particular, they are meromorphic and fully symmetric under permutation of its variables. The only poles of $\omega_{g,n}$ are at ramification points of the spectral curve.

Of our main interest in this paper is the wave function
\begin{equation}
\psi_{TR}(x)=\exp\big(\frac{1}{\hbar}S_0^{TR}(x)+S_1^{TR}(x)+\hbar S_2^{TR}(x)+\dots\big)    \label{wavef}
\end{equation}
where $x=e^u$, and
\begin{align}
\begin{split}
S_0^{TR}(x)&=\int\omega_{0,1},\\
S_1^{TR}(x)&=-\frac{1}{2}\log \frac{du}{dt},  \label{S0S1}
\end{split}
\end{align}
(where the formula for $S_1^{TR}(x)$ is valid only for curves of genus 0), while $S_k^{TR}(x)$ for $k>1$ are defined in terms of $\omega_{g,n}$ as follows
\begin{equation}
S_k^{TR}(x)=\sum_{2g-1+n=k} \frac{1}{n!}
\int_{t_0}^t\dots\int_{t_0}^t \omega_{g,n}(t_1,\dots,t_n),     \label{psi_coeffs}
\end{equation}
for appropriately chosen integration boundary $t_0$. The dependence on $x$ in the above formulas follows from inverting the parametrization $x=x(t)$. For example
\begin{equation}
\begin{aligned}
S_2^{TR}&=\frac{1}{3!}\iiint \omega_{0,3}+\int\omega_{1,1}, \\
S_3^{TR}&=\frac{1}{4!}\iiiint \omega_{0,4}+\frac{1}{2!}\iint \omega_{1,2} ,\\
S_4^{TR}&=\frac{1}{5!}\underbrace{\int\ldots\int}_{5\ \text{times}} \omega_{0,5}+\frac{1}{3!}\iiint \omega_{1,3}+\frac{1}{5!}\int\omega_{2,1}. \label{S2S3S4-TR}
%    \vdots
\end{aligned}
\end{equation}
The definition (\ref{wavef}) and the form of $S_k^{TR}(x)$ are motived by the case when the spectral curve is a spectral curve of some matrix model -- in this case the above wave function arises as the determinant expectation value $\big\langle \textrm{det}\frac{1}{x-M}\big\rangle$, where $\langle\cdot\rangle$ involves integration over an ensemble of matrices $M$.

It follows that (\ref{psi_coeffs}) can be determined order by order in $\hbar$ expansion, by means of the topological recursion. Furthermore, it is conjectured that there exists an operator, called quantum (spectral) curve, which annihilates the wave function (\ref{wavef}), and which can be regarded as a quantization of the spectral curve. For curves in $\mathbb{C}^*\times \mathbb{C}^*$ which will be of our main interest in this work, the quantum curve is an operator in variables $\hat x = e^{\hat u}$ and $\hat y = e^{\hat v}$, which satisfy the commutation relation $\hat y \hat x = q \hat x \hat y$. We denote this operator by $\widehat{A}_{TR}(\hat x, \hat y)$, and the equation it imposes reads
\begin{equation}
\widehat{A}_{TR}(\hat x, \hat y) \psi_{TR}(x) = 0.   \label{A-TR}
\end{equation}
In \cite{abmodel} it was shown that the above quantum curve can be also determined order by order in $\hbar$ by means of the topological recursion. In a similar vain, in this work we will show that Nahm sums or quiver motivic generating functions, as well as corresponding quantum curves, can be determined by the topological recursion. 

Note that for some quivers we are not able to write down explicitly the analytical form of higher order coefficients $S_k^{TR}$, for $k\geq 2$. Nonetheless, for quiver A-polynomials of genus 0 we are always able to compute $S_1^{TR}$ using (\ref{S0S1}), and then taking advantage of (\ref{rescale-Ahat}) we can postulate the form of the quantum curve that annihilates the full topological recursion partition function $\psi_{TR}(x)$. From the consistency of the topological recursion formalism we expect, that in this way we predict the correct form of $\widehat{A}_{TR}(\hat x, \hat y)$. The fact that such a curve exists, and it arises from a simple rescaling of generating parametres, also confirms our claim that the topological recursion formalism reconstructs quiver generating series and corresponding quantum quiver A-polynomials.

Finally, we stress that the topological recursion computations can be conducted for curves that we call admissible, i.e. they are irreducible and have a maximal number of ramification points (which cannot be increased by changing the framing). Indeed, if a given curve of interest does not have such a maximal number of ramification points, then the topological recursion misses some contributions and yields incorrect result  \cite{vb-ps}. In case some A-polynomial of our interest does not have the maximal number of ramification points, then we simply analyze one of its admissible cousins in different framing (which is a minor change, inessential in many applications). There is also a simple criterion of whether a given curve has the maximal number of ramification points -- it is so if its Newton polygon has no horizontal slopes \cite{Eynard:2012nj}.

%***************************************************************************************************
%***************************************************************************************************

\section{Nahm sums and quantum quiver A-polynomials from topological recursion}   \label{sec-TR}

One of the main statements in this work is that the Nahm sums, or equivalently quiver generating functions (\ref{QGF}), with appropriate identification of $x_i$'s with a single $x$, can be identified as the wave-functions (\ref{wavef}) associated the underlying classical A-polynomial. This means that an expansion of the quiver generating function can be determined, order by order in $\hbar$, by the topological recursion. This also means, that the difference equations that annihilate quiver generating functions (\ref{AhatPx}) can be determined perturbatively by the topological recursion.

More precisely, in the identification of variables mentioned above, variables $x_i$ in the quiver generating function are identified with a single $x$ up to rescalings by powers of $q$, and also the whole wave-function is rescaled by an overall monomial in $x$ (which therefore affects only $S_1(x)$ term in the asymptotic expansion of the wave function). The precise relation that we postulate therefore reads
\begin{equation}
\psi_{TR}(x) = x^d P_C(q^{d_1}x,\dots,q^{d_m}x).    \label{psiTR-PC}
\end{equation}
Note that the shifts by $q^{d_i}$ are analogous to those observed before in the knots-quivers correspondence \cite{Kucharski:2017ogk}. Due to the nature of the topological recursion, we compare (taking into account the above redefinitions) expansions in $\hbar$, which are given by (\ref{wavef}) on the topological recursion side, and by the WKB expansion (\ref{PC-S0}) of the quiver generating series.  Note that the leading term $S_0=\int \log y\frac{dx}{x}$ is captured just by the underlying classical A-polynomial and it takes the same form in the WKB expansion (\ref{PC-S0}) and in the topological recursion wave-function (\ref{S0S1}). For this reason, the crucial term which we have to match is the first correction $S_1$. For A-polynomials of genus 0, we need to match its form from the WKB expansion with $S_1^{TR}=-\frac12\log \frac{du}{dt}$ subleading term in the topological recursion wave-function (\ref{S0S1}). From this matching, and using (\ref{psi-qx}), we can in particular read off the shifts in (\ref{psiTR-PC}), as well as the relation between quantum A-polynomials (\ref{rescale-Ahat}). Once the correction $S_1$ is matched, subleading topological recursion computations of $S_2^{TR}, S_3^{TR},$ etc., on one hand confirm the consistency of the relation (\ref{psiTR-PC}), and on the other hand they may also detect the presence of non-uniform $q$-dependent coefficients in the quantum A-polynomial (if all coefficients in quantum A-polynomial would be monomials in $q$, they would be determined solely by $S_1$ term \cite{abmodel}).

Depending on (technical) complexity, we verify that the topological recursion wave-function (\ref{wavef}) reproduces the expansion of the generating functions for the above quivers (\ref{QGF}) to various orders in $\hbar$. In all cases of interest it is relatively easy to compute $S_1^{TR}$ term using (\ref{S0S1}). Note that $S_1$ from the WKB expansion of the quiver generating function is simply a function of $x$, while the form of $S_1^{TR}$ in (\ref{S0S1}) may depend on the parametrization of the underlying A-polynomial, as well as the choice of appropriate branch of A-polynomial; the existence of a parametrization for which these subleading corrections agree is therefore the first non-trivial check that we make. Moreover, whenever (analytically) possible, we also compute $S_2$ and higher order terms by the topological recursion. The form of these terms depends not only on the choice of parametrization, but also on the base point of integration $t_0$  in (\ref{psi_coeffs}). The existence of such $t_0$ is also a non-trivial feature of our postulate.

We verify whether the above identification works for various representative quivers, which were presented in section \ref{sec-our-quivers}. We choose these quivers so that the corresponding A-polynomials have genus 0 and are admissible (i.e. they are irreducible and have maximal number of ramification points); to this end we appropriately identify generating parameters $x_1=\pm x_2=x$ (so that the A-polynomial is irreducible), and introduce framing if necessary (so that the number of ramification points is maximal). In the following sections we analyze our statement, and conduct topological recursion calculations, for the following quivers:
\begin{itemize}
\item 1-vertex quiver discussed in section \ref{ssec-1vertex}, in arbitrary framing $f$, encoded in the matrix $C=[f]$; this quiver captures topological string amplitudes for $\mathbb{C}^3$ geometry, as well as extremal colored HOMFLY polynomials of the framed unknot,
\item uniform quivers (\ref{C-uniform}), of size 2 and independently of arbitrary size, with identification $x_i=x$, 
\item a quiver $C=\begin{bmatrix}
2 & 1 \\
1 & 1 
\end{bmatrix}$, which is an example of a quiver (\ref{family-a0}) from the family $C=\begin{bmatrix}
a & 0 \\
0 & 0 
\end{bmatrix}$ with $a=1$ and framing $f=1$, and with identification $x_1=-x_2=x$; this quiver captures full colored HOMFLY polynomials for the unknot,
\item a quiver $C=\begin{bmatrix}
3 & 1 \\
1 & 1 
\end{bmatrix}$, which is also an example of a quiver (\ref{family-a0}) from the family $C=\begin{bmatrix}
a & 0 \\
0 & 0 
\end{bmatrix}$ with $a=2$ and framing $f=1$, and with identification $x_1=x_2=x$; this quiver encodes extremal colored invariants of the left-handed trefoil knot,
\item a quiver $C=\begin{bmatrix}
2 & 0 \\
0 & -1 
\end{bmatrix}$, which is an example of a reciprocal quiver $\begin{bmatrix}
a & 0 \\
0 & 1-a 
\end{bmatrix}$ presented in (\ref{family-a1-a}) with $a=2$, and with identification $x_1=-x_2=x$,
\item a quiver $C=\begin{bmatrix}
2 & 2 \\
2 & 1 
\end{bmatrix}$, which is an example of $\begin{bmatrix}
c & c \\
c & 0 
\end{bmatrix}$ in (\ref{family-ccc0}) with $c=1$ and framing $f=1$, and with identification $x_1=x_2=x$,
\item a quiver $C=\begin{bmatrix}
3 & 2 \\
2 & 1 
\end{bmatrix}$, which is an example of $\begin{bmatrix}
c+1 & c \\
c & 0 
\end{bmatrix}$ in (\ref{family-c+1cc0}) with $c=1$ and framing $f=1$, and with identification $x_1=x_2=x$; this quiver encodes extremal colored invariants of the right-handed trefoil knot,
\item a diagonal quiver $\begin{bmatrix}
2 & 0 \\
0 & 2 
\end{bmatrix}$ introduced in (\ref{diagonal-22}), with identification $-x_1=x_2=x$,
%\item a quiver $\begin{bmatrix}
%1 & 2 \\
%2 & 1 
%\end{bmatrix}$, which is also example of a diagonal quiver (\ref{family-aa}) with $a=-1$ and framing $f=2$, with identification $-x_1=x_2=x$.
\end{itemize} 
We find agreement between quiver generating functions and topological recursion calculations in all cases, apart from one subtlety for diagonal quivers (the last example). Namely, in this case one term in the quantum A-polynomial determined from the topological recursion appears to have a different ordering than the operator that annihilates the quiver generating function. It would be important to explain this discrepancy.

%\be
%\underset{x_1=x_2=x}{\begin{pmatrix}
%1 & 1 \\
%1 & 1 
%\end{pmatrix}}\qquad 
%\underset{x_1=-x_2=x}{\begin{pmatrix}
%2 & 1 \\
%1 & 1 
%\end{pmatrix}}\qquad
%\underset{-x_1=x_2=x}{\begin{pmatrix}
%2 & 0 \\
%0 & 2
%\end{pmatrix}}\qquad
%\underset{x_1=-x_2=x}{\begin{pmatrix}
%2 & 0 \\
%0 & -1
%\end{pmatrix}}\qquad
%\underset{x_1=x_2=x}{\begin{pmatrix}
%2 & 2 \\
%2 & 1 
%\end{pmatrix}}\qquad
%\underset{-x_1=x_2=x}{\begin{pmatrix}
%1 & 2 \\
%2 & 1 
%\end{pmatrix}}
%\ee

%***************************************************************************************************

\subsection{One-vertex quiver, $m=1$}   \label{ssec-1vertex-TR}

We analyze first a quiver generating function with $m=1$, associated to a quiver that consists of one vertex and, in general, $f$ loops. A basic data associated to this quiver was introduced in section \ref{ssec-1vertex}. This example was also analyzed in \cite{vb-ps,abmodel} from the viewpoint of topological string theory, and it also captures extremal invariants of the unknot \cite{Kucharski:2017ogk}. It is however instructive to discuss it again from our current perspective and fill in various details. To start with we analyze $f=2$ case, and then arbitrary framing $f$.

\subsubsection*{Framing $f=2$}

Let us consider first framing $f=2$, which is the lowest framing for which the A-polynomial is admissible. In this case quantum and classical A-polynomials take form
\begin{equation}
	\hat{A}(\hat{x}, \hat{y}) = q \hat{x} \hat{y}^2 + \hat{y} - 1,\qquad A(x,y) = x y^2 + y - 1.   \label{Ahat-A-1vertex}
\end{equation}
The leading and subleading terms in the WKB expansion of the quiver generating function (\ref{1vertex-param}) take form
\begin{align}
\begin{split}
	S_0 &= - {\rm Li}_2(1-y) - (\log y)^2, \\
	S_1 &= - \frac{1}{2}\log \frac{(1-y)y}{x} - \frac{1}{2}\log \left(- \frac{dx}{dy}\right) = - \frac{3}{2}\log y - \frac{1}{2} \log \left(- \frac{dx}{dy}\right).   \label{S0S1-1vertex-wkb}
\end{split}
\end{align}
Note that $y(x)$ is a monotonically decreasing function, so $- dx/dy > 0$. 

We now apply the topological recursion to the classical A-polynomial in (\ref{Ahat-A-1vertex}). We consider parametrization (\ref{1vertex-param}), which for $f=2$ takes form
\begin{equation}
	x(t) = \frac{1-t}{t^2}, \qquad y(t) = t, \label{first_parametrization}
\end{equation}
with $t \in [0, 2]$. This range of parameter describes the whole branch of $y(x)$ solving $A(x,y)=0$ with $y(0)=1$, that is
\begin{equation}
	y(x) = \frac{-1 + \sqrt{1 + 4x}}{2x}.
\end{equation} 
Note that $y(x)$ becomes complex for $x < -1/4$. This is the point where two branches of solution to $A(x,y)=0$ meet. On the quiver side it signals that the generating series stops to converge. %For this reason we focus on $x>0$. 

Evaluating general formulas (\ref{S0S1}) in the parametrization (\ref{first_parametrization}) we find
\begin{align}
\begin{split}
	S_0^{TR} &= \int \log y \frac{dx}{x} = -{\rm Li}_2(1-y) - (\log y)^2 = S_0,\\
	S_1^{TR} &= -\frac{1}{2} \log \frac{1}{x} \frac{dx}{dt} = -\frac{1}{2} \log \frac{-2 +y}{x y^3}.
\end{split}
\end{align}
The leading term of course reproduces $S_0$ in (\ref{S0S1-1vertex-wkb}), while comparing with $S_1$ in (\ref{S0S1-1vertex-wkb}) we find
\begin{equation}
	S_1^{TR} = S_1 + \frac{1}{2} \log (-x) + \frac{3}{2}\log y. 
\end{equation}
Using (\ref{psi-qx}), this leads to the following conjecture for the relation between $P_C(x)$ and $\psi_{TR}(x)$
\begin{equation}
	P_C(q^{3/2} x) = i x^{-1/2} \psi_{TR}(x). \label{f=2_first_conjecture}
\end{equation} 
Furthermore, using (\ref{rescale-Ahat}), this leads to the following quantum A-polynomial annihilating $\psi(x)_{TR}$ (which agrees with the result in \cite{abmodel})
\begin{equation}
	\hat{A}_{TR}(\hat{x}, \hat{y}) = \hat{A}(q^{3/2} \hat{x}, q^{-1/2} \hat{y}) = q^{3/2} \hat{x} \hat{y}^2 + q^{-1/2} y - 1. 
\end{equation}
% up to $(x \rightarrow-x, y\rightarrow -y)$ in agreement with different choice for classical A-polynomial.

We now confirm that $S^{TR}_k$ with $k>1$ are consistent with the above relations. First we compute auxiliary topological recursion data. There is one branching point $\frac{d \log x(t)}{dt}|_{t=t^*} = 0$ that arises for $t^* = 2$. The conjugate point, defined via $x(t) = x(\bar{t})$ for $\bar{t}$ in the neighbourhood of $t^*$, takes form $\bar{t} = \frac{t}{t-1}$. The differential $1$-form, for parametrization~\eqref{first_parametrization} and the above conjugate point, reads
\begin{equation}
	 \omega(t, \bar{t}) = \left( \log y(t) - \log y(\bar{t})\right) \frac{dx(t)}{x(t)} =  \log(t-1) \frac{-2+t}{t(1-t)} dt.
\end{equation}
For the Bergman kernel for genus 0 curves (\ref{Bergman}), its associated $1$-form reads
	 \begin{equation}
	 	dE_{t,\bar{t}}(t') = \frac{1}{2}\int_{\xi=\bar{t}}^{t} B(\xi, t') = -\frac{t}{2} \frac{2-t}{(t'(t-1)-t)(t'-t)} dt'.
	 \end{equation}
Finally, the recursion kernel takes form
\begin{equation}
	 K(t', t, \bar{t}) = \frac{dE_{t, \bar{t}}(t')}{\omega(t, \bar{t})} = \frac{t^2}{2}\frac{1-t}{(t'(t-1)-t)(t'-t)} \frac{1}{\log(t-1)} \frac{dt'}{dt}.
\end{equation}
We now evaluate
\begin{equation}
	\omega_{0,2}(\bar{q}, t_1) = \frac{d\bar{q} dt}{(\bar{q} - t)^2} = - \frac{1}{1-q^2} \frac{dq dt}{(\bar{q} - t)^2},
\end{equation}
and using (\ref{w11-w03}) we get
\begin{align}
\begin{split}
	\omega_{1,1}(t) &= - \frac{16 - 16 t + t^2}{24(2-t)^4} dt, \\
	\omega_{0,3}(t_1, t_2, t_3) &= \frac{4 dt_1 dt_2 dt_3}{(t_1 - 2)^2 (t_2 - 2)^2 (t_3 - 2)^2}.
\end{split}
\end{align}

\begin{figure}[h]
	\begin{center}
	\includegraphics[scale=0.6]{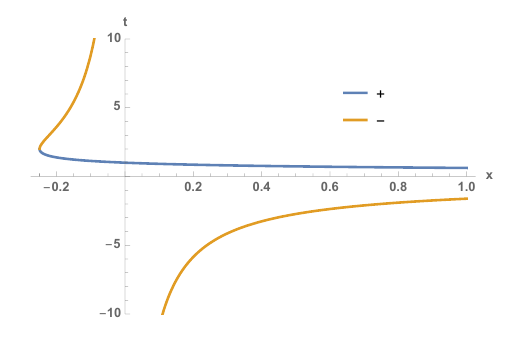}
	\caption{Plot of $t_{\pm}(x) = \frac{-1 \pm \sqrt{1 + 4x}}{2x}$. }
	\label{fig:tpm(x)}
	\end{center}
\end{figure}

We can now compute $S_2^{TR}$, as given by (\ref{S2S3S4-TR}), and, for the base point $t_0 = -\infty$, we find
\begin{equation}
	S_2^{TR}(t) = \frac{-4 - 10 t + t^2}{24 (t-2)^3}.
\end{equation}
For the comparison with Nahm sums we have to express $S_2^{TR}$ as a function of $x$. To this end we invert the relation $x(t)$. There are two branches (see Fig.~\ref{fig:tpm(x)}),
\begin{equation}
	t_{\pm}(x) = \frac{-1 \pm \sqrt{1 + 4x}}{2x}, \label{tpm(x)_2}
\end{equation}
out of which we choose the one for which $y(x=0) = 1$, this corresponds to the $t_+(x)$-branch. Note that in the parametrization that we are using $y=y(t) = t$ with $y$ the regular solution of the $A$-polynomial $A(x,y) = 0$, which indeed is
\begin{equation}
	y(x) = \frac{- 1 + \sqrt{1+4x}}{2x}.
\end{equation}
Therefore the final answer for $S_2$ is
\begin{equation}
	S_2^{TR}(x) = \frac{-4 - 10 y(x) + y(x)^2}{24 (y(x)-2)^3},
\end{equation}
in agreement (up to an irrelevant constant) with the WKB expansion of the quiver generating series.

Alternatively, we can compute $S_2^{TR}$ with the base point $t_0 = 1$,
\begin{equation}
	S_2^{TR}(t) = - \frac{-4 + 22t - 31 t^2 +13 t^3}{24(t-2)^3},
\end{equation}
and this time insert the other branch $t_-(x)$. Noting that, the two branches are related by the Galois involution, $t_-(x) = t_+(x)/ (1 - t_+(x))$ we obtain the same answer (up to a sign) as before
\begin{equation}
	S_2^{TR}(x) = - \frac{-4 + 22t_-(x) - 31 t_-^2(x) +13 t_-^3(x)}{24(t_-(x)-2)^3} = - \frac{-4 - 10 y(x) + y(x)^2}{24 (y(x)-2)^3}. 
\end{equation}

%***************************************************************************************************

\subsubsection*{Arbitrary framing $f$}

Now we summarize the computation for arbitrary framing $f$. A quantum curve, first terms in WKB expansion, and other details are introduced in section \ref{ssec-1vertex}. $S_0^{TR}(x)$ trivially reproduces the leading term given in (\ref{S0S1-1vertex}), while from (\ref{S0S1}) we find the relation between the subleading term $S_1(x)$ in WKB expansion and the topological recursion
\be 
S_1^{TR}(x) = S_1(x) + \frac{1}{2} \log x + \frac{f+1}{2} \log y + \frac{1}{2} \log (-1)^{f+1}.
\ee
Taking advantage of (\ref{rescale-Ahat}) and (\ref{psi-qx}), this leads to the following conjectural relation between $P_C(x)$ and $\psi_{TR}(x)$
\begin{equation}
	\psi_{TR}(x) = (-1)^{(f+1)/2} x^{1/2} P_C(q^{(f+1)/2} x), \label{conjecture_f}
\end{equation}
and we find the following quantum $A$-polynomial annihilating $\psi_{TR}(x)$ 
\begin{equation}
	\hat{A}_{TR}(\hat{x}, \hat{y}) = \hat{A} (q^{(f+1)/2} \hat{x}, q^{-1/2} \hat{y}) = (-1)^f q^{(f+1)/2} \hat{x} \hat{y}^{f} + q^{-1/2} \hat{y} - 1.
\end{equation}
We now compute higher terms $S_k^{TR}$ to verify the above conjectural results. Using the parametrization in (\ref{1vertex-param}), we find the branch point $t^* = f/(f-1)$. There is no closed form expression for the conjugate point, however for practical computations it is sufficient to determine
	\begin{align}
	\begin{split}
		\bar{t} =&\, t^* - (t - t^*) + \frac{2(f^2-1)}{3f}(t-t^*)^2 - \frac{4 (f^2 - 1)^2}{9 f^2} (t- t^*)^3  +\\
		&+\frac{2(f-1)^3(22 f^3 + 57 f^2 + 57 f + 22)}{135 f^3} (t - t^*)^4 + \mathcal{O}((t-t^*)^5).
	\end{split}
	\end{align}
Furthermore, the vertex $\omega(t_1, t_2)$, one-form $dE$ and the recursion kernel take form
\begin{align}
\begin{split}
\omega(t, \bar{t}) &= (\log t - \log \bar{t}) \frac{f(t-1)-t}{t(1-t)} dt,\\
dE_{t, \bar{t}}(t') &= \frac{1}{2} \frac{t - \bar{t}}{(t' - t)(t' - \bar{t})} dt', \\
K(t', t, \bar{t}) &= \frac{dE_{t, \bar{t}}(t')}{\omega(t, \bar{t})} = \frac{1}{2} \frac{\bar{t} - t}{(t' - t)(t' - \bar{t})} \frac{t(1-t)}{f(t-1)-t} \frac{1}{\log \bar{t}/t}\frac{dt'}{dt}.
\end{split}
\end{align}
From the above results we now compute 
\begin{align}
\begin{split}
	\omega_{1,1}(t) &= \frac{1}{24 (f-1)^4} \frac{- t^2 (f-1)^4 + 2 t f (f^3 - 2f^2 + 3f - 2) - f^4}{(t - t^*)^4},\\
	\omega_{0,3}(t_1, t_2, t_3) &= \frac{f^2}{(f-1)^4}\frac{1}{(t_1 - t^*)^2 (t_2 - t^*)^2 (t_3 - t^*)^2},
\end{split}
\end{align} 
and higher differentials (whose form is too involved to type here). With these results, we can finally determine $S_2^{TR}(t)$ and $S_3^{TR}(t)$ using (\ref{S2S3S4-TR}). We find, for the base point $t_0 = -\infty$,
\begin{align}
\begin{split}
	S_2^{TR}(t) &= - \frac{1}{(f-1)^4} \frac{- (f-1)^4 t^2 + f (f-1)(3 -3f +2f^2)t + f^2 (-f^2 + f +3)}{24 (t -t^*)^3},\\
	S_3^{TR}(t) &= - \frac{f t(t-1)\left[ 2f^4 (t-1)^2 - 8 f^2 - ( 9f^2 - 14 f + 9) f t(t-1) + 2 t^2\right]}{48 (f(1-t) + t)^6}.
\end{split}
\end{align}
We use now regular branch of $x(t)$ to invert this relation. We find
\begin{equation}
	t(x) = 1 + (-1)^{f+1} x + f x^2 + \frac{1}{2}(-1)^f (f - 3f^2) x^3 + \frac{1}{3}f(1 - 6f +8f^2) x^4 + \dots
\end{equation}
which allows us to find an expansion of $S_2^{TR}(x)$ and $S_3^{TR}(x)$ in $x$. We then find that these results reproduce the WKB expansion, in agreement with the conjecture~\eqref{conjecture_f}.

%***************************************************************************************************

\subsection{Uniform two-vertex quivers, $m=2$}

We consider now a uniform quiver, discussed in section (\ref{ssec-uniform}). We focus first on a quiver of size $m=2$ and $f=1$, which is represented by a matrix
\be
C=\begin{bmatrix}
1 & 1 \\
1 & 1
\end{bmatrix}    \label{C1111}
\ee
The resultant (\ref{CAPUQ}) in this case takes form  
\begin{equation}
A(x_1,x_2;y) = {y}^{2}x_{{1}}x_{{2}}+ \left( x_{{1}}+x_{{2}}-1 \right) y+1.
\end{equation}
In this example we consider the A-polynomial, which arises upon the specialization $x_1 = x, x_2 = x$, and thus it takes form 
\begin{equation}
A(x,y) = \left( xy+1 \right) ^{2}-y = 0,    \label{Q1111-Axy}
\end{equation}
which is parametrized as in (\ref{uniform-param})
\begin{equation}
x(t) = \frac{t-1}{t^2},\qquad y(t) = t^2.   \label{Q1111-param}
\end{equation}
Note that the inverse relation $t(x)={\frac {1-\sqrt {1-4x}}{2x}}$ is regular at $x = 0$.
Furthermore, the quantum generating function (\ref{QGF}) with the specialization $x_1=x_2=x$ is annihilated by the quantum curve operator (\ref{Ahat-xy-uniform})
\begin{equation}
\widehat{A}(\hat{x},\hat{x},\hat{y}) = q^2\hat{x}^2\widehat{y}^2+2\sqrt{q}\hat{x}\widehat{y}-\widehat{y}+1. \label{AhaQ_uni}
\end{equation}

Let us show how the quantum generating function (\ref{QGF}) with the above specialization of $x_i$, and the quantum curve (\ref{AhaQ_uni}), are consistent with (and thus can be determined by) the topological recursion. To this end, we show that the wave function $\psi_{TR}(x)$ in (\ref{wavef}) -- associated (via the topological recursion) to the classical curve (\ref{Q1111-Axy}) -- is annihilated by (\ref{AhaQ_uni}), after a simple rescaling of variables
\begin{equation}
\widehat{A}(q^{\alpha_1}x,q^{\alpha_2}x;q^{\beta}\widehat{y})\psi_{TR}(x) = 0. \label{Ahatr_uni}
\end{equation}
To determine $\alpha_1,\alpha_2$, and $\beta$, let us consider a perturbative expansion of this equation, using the form (\ref{wavef}). According to (\ref{S0S1}), and using the parametrization (\ref{Q1111-param}) of the classical curve (\ref{Q1111-Axy}) we get
\begin{equation}
\begin{aligned}
S_0^{TR} =&\ \int {\frac {1}{x}\ln  {\frac { ( X-1 ) ^{2}}{4{x}^{
				2}}} } dx,  \\
S_1^{TR} =&-\frac{1}{2}\Big(\ln(2) + \ln  \Big( {\frac { (X+4x-1) x}{ ( X+2x-1)  ( X-1 ) }}\Big)
\Big) ,   \label{C1111-S0S1}
\end{aligned}
\end{equation}
\bigskip
where $X=\sqrt {1-4x}$. The first term (of order $\hbar^0$) in the expansion of (\ref{Ahatr_uni}) then vanishes automatically, while the vanishing of the subleading term (of order $\hbar$) imposes the unique (with $\alpha_1=\alpha_2$, as expected due to the symmetry of the quiver) solution
%\begin{equation}
%\begin{aligned}
%0 = \ & \left( 16\,\beta+8\,\alpha_{{1}}+8\,\alpha_{{2}}-8 \right) {X}^{2}{x}^{3}+ \left( -36\,\beta-26\,\alpha_{{1}}-26\,\alpha_{{2}}+34 \right) {X}^{2}{x}^{2}+\\
%&
 %\left( 16\,\beta+14\,\alpha_{{1}}+14\,\alpha_{{2}}-20\right) {X}^{2}x+ \left( -2\,\beta-2\,\alpha_{{1}}-2\,\alpha_{{2}}+3\right) {X}^{2}+ \\
%& \left( -80\,\beta-48\,\alpha_{{1}}-48\,\alpha_{{2}}+56 \right) {x}^{3}X+ \left( 100\,\beta+76\,\alpha_{{1}}+76\,\alpha_{{2}}-102 \right) {x}^{2}X+ \\
%& \left( -36\,\beta-32\,\alpha_{{1}}-32\,\alpha_{{2}}+46 \right) Xx+ \left( 4\,\beta+4\,\alpha_{{1}}+4\,\alpha_{{2}}-6 \right) X+ \\
%& \left( 64\,\beta+40\,\alpha_{{1}}+40\,\alpha_{{2}}-48\right) {x}^{3}+ \left( -64\,\beta-50\,\alpha_{{1}}-50\,\alpha_{{2}}+68 \right) {x}^{2}+ \\
%& \left( 20\,\beta+18\,\alpha_{{1}}+18\,\alpha_{{2}}-26 \right) x-2\,\beta-2\,\alpha_{{1}}-2\,\alpha_{{2}}+3.
%\end{aligned}
%\end{equation}
\begin{equation}
\alpha_1=\alpha_2=1,\qquad \beta=-\frac{1}{2}.    \label{alpha_beta_uni}
\end{equation}
With these values, the form of the topological recursion quantum curve (\ref{A-TR}) is fixed
\begin{equation}
\widehat{A}_{TR}(\hat x, \hat y) =\widehat{A}(q^{\alpha_1}x,q^{\alpha_2}x;q^{\beta}\widehat{y}) 
= {q}^{3}\hat{x}^{2}\widehat{y}^{2}+2\,q\hat{x}\widehat{y}-{\frac {\widehat{y}}{\sqrt {q}}}+1.  \label{AhatTR_uni}
\end{equation}
Nonetheless, we should check that higher order terms of the topological recursion wave function are also consistent with this result. From the equation 
\be
\widehat{A}_{TR}(\hat x, \hat y) \exp\Big(  \frac{1}{\hbar}S_0^{TR} + S_1^{TR} + \sum_{k=1}^{\infty} \hbar^k S_k \Big)= 0,
\ee
with the results (\ref{C1111-S0S1}) and (\ref{AhatTR_uni}), the following higher order terms follow perturbatively 
\begin{equation}
\begin{aligned}
S_2 =& {\frac {12\,{x}^{2}+3\,X-14\,x+4}{24{X}^{3}}}, \\
S_3 =& {\frac { \left( 4\,{x}^{2}-2\,x-1 \right) x \left( 2\,Xx-X+4\,x-1
		\right) }{8{X}^{7}}}, \\
S_4 =& -\frac {2}{{X}^{9}} \Big( X ( {x}^{5}-5/4\,{x}^{4}+2/3\,{x}^
{3}-{\frac {11\,{x}^{2}}{32}}+{\frac {x}{192}}-{\frac{1}{1024}}
)+\\
& {-\frac {25\,{x}^{6}}{12}}+{\frac {73\,{x}^{5}}{24}} -{\frac {
		1243\,{x}^{4}}{720}}+{\frac {119\,{x}^{3}}{144}}-{\frac {203\,{x}^{2}
	}{960}}-{\frac {67\,x}{5760} \Big) }.       \label{Swkb}
\end{aligned}
\end{equation}
We should therefore confirm that these terms agree with the topological recursion computation. To this end we again focus on the spectral curve (\ref{Q1111-Axy}) and its parametrization (\ref{Q1111-param}). There are two ramification points, i.e. the preimages $t$ of $x$ where $x$ has branching, at $t^*=2$ and at $t^*=0$. Notice that the second point corresponds to $x=\infty$, which therefore must be included to the residue formula. One way to see that it contributes is by noticing that $x$ has a pole, while $y$ has zero, at the same point, of the same degree (two). 
%This does not happen for one-vertex quiver in any framing!

With all this data, we now compute $\omega_{g,n}$; in particular we find
\begin{equation}
\begin{aligned}
\omega_{1,1} =&{\frac {t_{{1}}}{ 4\left( t_{{1}}-2 \right) ^{4}}}-\frac{1}{4}\, \left( t_
{{1}}-2 \right) ^{-4}-\frac{1}{48}\, \left( t_{{1}}-2 \right) ^{-2}-\frac{1}{16}\,{t_{
		{1}}}^{-2}, \\
\omega_{0,3} =&{\frac {2}{ \left( t_{{1}}-2 \right) ^{2} \left( -2+t_{{2}}
		\right) ^{2} \left( t_{{3}}-2 \right) ^{2}}} ,
%\omega_{0,4} =&\ 3([4,2,2,2] + [3,2,2,2]) + 2\cdot [2,2,2,2] \\
%\omega_{1,2} =&\ {\frac {T \left( t_{{1}},t_{{2}} \right) }{ \left( -2+t_{{2}} \right) 
%		^{6} \left( t_{{1}}-2 \right) ^{6}\boxed{{t_{{1}}}^{2}{t_{{2}}}^{2}}}} = \\
%&\ + \boxed{holo|_{t=2}}
\end{aligned}
\end{equation}
(note that the last term $-\frac{1}{16}\,{t_{{1}}}^{-2}$ in $\omega_{1,1}$ arises from the residue at $t^*=0$, i.e. $x=\infty$).
%We do not write here expressions for $\omega_{0,4}$ and $\omega_{1,2}$, but we computed them as well.
Computing now $S_2^{TR}$ in (\ref{psi_coeffs}) we find
\begin{equation}
\begin{aligned}
S_2^{TR} =&-{\frac {{t}^{3}-7\,{t}^{2}+12\,t-6}{12 t \left( t-2 \right) ^{3}}}+
{\frac {{t_{{0}}}^{3}-7\,{t_{{0}}}^{2}+12\,t_{{0}}-6}{12 t_{{0}}
		\left( t_{{0}}-2 \right) ^{3}}}+
	\\ 
	&\ + {\frac { \left( t-t_{{0}}
		\right) ^{2}}{ 3 \left( t-2 \right) ^{3} \left( t_{{0}}-2 \right) ^{2}}
}-{\frac { \left( t-t_{{0}} \right) ^{2}}{3 \left( t-2 \right) ^{2
		} \left( t_{{0}}-2 \right) ^{3}}}.
\end{aligned}
\end{equation}
If we choose the integration limit as $t_0=1$ (which is the zero of $x=x(t)$; also note that $x(t_0) = 0, y(t_0) = 1$), then this result agrees with $S_2$ in (\ref{Swkb}) up to an irrelevant constant, which can be absorbed into an overall normalization of the wave function
\begin{equation}\label{S2tr}
S_2^{TR} = {\frac {4\,{t}^{4}-13\,{t}^{3}+19\,{t}^{2}-16\,t+6}{12 t \left( t-
		2 \right) ^{3}}} = S_2 + \frac{1}{24}.
\end{equation}
Note that, due to the presence of the square root in $S_2$, upon inserting the parametrization, one has to choose the right branch, in this case it is $t < 0$ or $t>2$.  
Similarly, with the same integration limit $t_0=1$, we find that 
\begin{equation}
\begin{aligned}
S_3^{TR} =&{\frac { \left( t-1 \right) ^{3} \left( {t}^{4}+2\,{t}^{3}-6\,{t}
		^{2}+8\,t-4 \right) }{4{t}^{2} \left( t-2 \right) ^{6}}},
\end{aligned}
\end{equation}
and it is also consistent with $S_3$ in (\ref{Swkb}). Analogously, we confirmed the consistency for the term $S_4^{TR}$.

The other choice for the integration limit is $t_0 = -\infty$. Then, upon restricting $t \in (0,2)$ when evaluating $S_2$ and $S_3$, we find consistently that 
\begin{equation}
\begin{aligned}
	S_2^{TR} &= -\frac{-6 + 8t - 7t^2 +t ^3}{12 t (t-2)^3} = S_2 + \frac{1}{24}, \\
	S_3^{TR} &= \frac{4-12 t + 14 t^2 - 8 t^3 + t^4 + t^5}{4 t^2(t-2)^6} = S_3,
\end{aligned}
\end{equation}

To sum up, up to appropriate shifts (\ref{alpha_beta_uni}) in (\ref{Ahatr_uni}), and with appropriately chosen integration limit $t_0=1$ or $t_0 = -\infty$, the topological recursion indeed reproduces the expansion of the motivic generating series associated to the quiver (\ref{C1111}), as well as the corresponding quantum curve (\ref{AhaQ_uni}).

%***************************************************************************************************

\subsection{Uniform quivers of arbitrary size $m$} 

We consider now uniform quivers of size $m \times m$ with all entries equal to $f$, whose properties we discussed in section \ref{ssec-uniform}. In particular, the parametrization of the classical A-polynomial reads
\begin{equation} 
	x(t) = \frac{1-t}{(-t^m)^f}, \qquad y(t) = t^m,
\end{equation}
which yields
\begin{align}
	S_0^{TR} &= - m {\rm Li}_2(1 - t) - \frac{m}{2} fm (\log t)^2,\\
	S_1^{TR} &= - \frac{1}{2} \log \frac{t - fm(t-1)}{t(t-1) }.
\end{align}
From the form of $S_1^{TR}$ term and using (\ref{psi-qx}) we conjecture
\begin{equation}
	\psi_{TR}(x) = x^{1/2} P_C(q^{(f+1)/2} x).    \label{psiTR-PC-uniform}
\end{equation}
Recall that the quiver generating function $P_C(x)$ in this case is annihilated by the operator $\widehat{A}(\hat x, \hat y)$ determined in (\ref{Ahat-xy-uniform}). Taking into account rescalings in (\ref{psiTR-PC-uniform}) and the relation (\ref{rescale-Ahat}), we predict that the operator that annihilates the topological recursion wave-function takes form
\begin{equation}
	\hat{A}_{TR}(\hat{x}, \hat{y}) = \hat{A}(q^{(f+1)/2} \hat{x}, q^{-1/2} \hat{y}) = \Big(1 - (-1)^f q^{(f+1)/2} \hat{x} \hat{y}^f \Big)^m - q^{-1/2} \hat{y}.
\end{equation}
As a consistency check, for $m=2$ and $f=1$ this reduces to (\ref{AhatTR_uni}).

In what follows we confirm the relation (\ref{psiTR-PC-uniform}) at higher orders in $\hbar$ expansion, using the topological recursion. There are two branch points: $t^*_1 = mf/(mf-1)$ and $t^*_2 = 0$. Let us present $S_k^{TR}$ as a sum of contributions arising from these two branch points
\be
S_k^{TR} = S_k^{TR, t^*_1} + S_k^{TR,t^*_2}.  \label{Sk-uniform}
\ee
We note that the contribution from the former branch point is simply related to the contribution for the one-vertex quiver, derived in section \ref{ssec-1vertex-TR}, just with $f$ replaced by  $fm$ and including additional rescaling
\begin{equation}
	S_k^{TR, t^*_1} \equiv S_k^{TR, (m,f)}(t) = m^{1-k} S_k^{TR, (1,mf)},   
\end{equation}
where $S_k^{TR, (m,f)}$ denotes $S_k$ term for a uniform quiver of size $m$ with all elements equal to $f$, computed by topological recursion; in particular $S_k^{TR, (1,n)}$ denotes results already computed in section \ref{ssec-1vertex-TR}. Introducing $n = m f$ and $t^* = n/(n-1)$, we find for example
\begin{align}
	S_2^{TR, (m,f)} &= - \frac{1}{m (n-1)^4} \frac{- (n-1)^4 t^2 + n (n-1)(3 -3n +2 n^2)t + n^2 (-n^2 + n +3)}{24 (t -t^*)^3}, \\
	S_3^{TR, (m,f)} &= - \frac{n t(t-1)}{m^2(n-1)^6}\frac{2n^4 (t-1)^2 - 8 n^2 - ( 9n^2 - 14 n + 9) n t(t-1) + 2 t^2 }{48 m^2 (n(1-t) + t)^6}.
\end{align} 

We now analyse the contributions from $t^*_2 = 0$. First, the conjugate point is given by the condition $\frac{1-t}{t^n} = \frac{1 - \bar{t}(t)}{\bar{t}(t)^n}$ (for $n = m f$). We thus write $\bar{t}(t) = a t p(t)$ where $a$ is a constant and $p(t)$ is a polynomial in $t$ such that $p(0) = 1$, satisfying the equation $(1-t)a^n p(t)^n = (1 - a p(t))$. We denote solutions of this equation by $(a_k, p_k(t))$ for $k=1, \dots, n$, with $a_k=\exp\left(\frac{2\pi i k}{n} \right)$, so that $(a_n = 1, p_n(t) = 1)$, and all other solutions implement Galois involutions $\bar{t}_k(t) = a_k t p_k(t)$. Furthermore, the vertex for each Galois involutions, in the vicinity of $t=0$, takes form
\be 
\omega^{(k)}(t, \bar{t}) = (\log y(t) - \log y(\bar{t}_k)) \frac{n (t-1) - t}{t(1-t)} dt = (\log a_k^m + m \log p_k(t)) \frac{t - n (t-1)}{t(1-t)} dt,
\ee
which leads to the following expression for the kernel
\begin{equation}
K^{(k)}(t', t, \bar{t}) = \frac{1}{2} \frac{\bar{t}_k - t}{(t' - t)(t' - \bar{t}_k)} \frac{t(1-t)}{m f(t-1)-t} \frac{1}{\log a_k^m + m \log p_k(t)}\frac{dt'}{dt}.
\end{equation}
Note that the kernel has different behaviour for $t\sim 0$ depending on whether $a_k^m = 1$. If so, the kernel is linear in $t$, and otherwise it is quadratic in $t$. Only when the kernel is linear in $t$ there will be a contribution to $S_k$. The condition $a_k^m = 1$ is fulfilled when $k/f$ is integer, and then the leading part of the kernel is $K^{(k)}(t', t, \bar{t}) = \frac{1}{2m} \frac{t}{t'^2} + \mathcal{O}(t^2)$.
		
With the above data, we compute now differentials (\ref{w11-w03}). $\omega_{0,2}(q, \bar{q})$ has a second order pole when $q\rightarrow 0$ which is cancelled if $K^{(k)}(t, q, \bar{q})$ is quadratic in $q$. Otherwise, when $a_k^m=1$, the residue is non-zero. On the other hand, any potential poles contributing to $\omega_{0,3}$  must come from the kernel itself, so $\omega_{0,3}$ is always zero. Thus contributions for a given $k$ reads
\begin{equation}
	\omega^{(k)}_{1,1}(t) = \begin{cases}
		\frac{1}{2m} \frac{a_k}{(1-a_k)^2 t^2},  &a_k^m = 1, \\
		0, &a_k^m \neq 1.
	\end{cases},
	\qquad  \quad
	\omega^{(k)}_{0,3}(t_1, t_2, t_3) = 0,
\end{equation}
and the contribution to $S_2(t)$ (for a given $k$) is $-\frac{1}{2m} \frac{a_k}{(1-a_k)^2 t}$. In total, the contribution from $t^*_2$ in (\ref{Sk-uniform}) reads
\begin{equation}
	S_2^{TR,t^*_2} = - \frac{1}{2m t}\sum_{\substack{k=1\\ k|f}}^{m-1} \frac{a_k}{(1-a_k)^2},
\end{equation}
with the sum over integers $k$ divisible by $f$. There are two immediate conclusions. First, there are no contributions of this type for a one-vertex quivers, since then $n = f$. Second, there are always contributions of this type for $f=1$ for quivers of size $m\geq 2$. Note that the value of the above sum for $f \geq 1$ does not depend on $f$. Introducing the notation
\begin{equation}
	A_m = \sum_{k=1}^{n-1} \frac{b^k}{(1-b^k)^2}, \qquad b =e^{2\pi i/m},
\end{equation}
the complete expression for $S_2^{TR}(t)$ for arbitrary uniform quiver with framing at least $1$ can be written as
\begin{equation}
	S_2^{TR}(t) = \frac{1}{m} S_2^{TR, (1, m f)}(t) - \frac{A_m}{2m t}.
\end{equation}
This indeed agrees with $S_2$ term in the WKB expansion of the generating series for the uniform quiver (with rescalings in (\ref{psiTR-PC-uniform}) taken into account).

We also confirm agreement for $S_3$ terms. Since $\omega_{1,1}$ is modified comparing to the one-vertex quiver, there might be extra contributions to $\omega_{2,1}$ and $\omega_{0,4}$ at both branch points $t^*_1=mf/(mf - 1)$ and $t^*_2=0$. The extra term in $\omega_{1,1}$ is 
\begin{equation}
	\frac{A_m}{2m t^2}.
\end{equation}
Let us start with the first branch point. The extra contributions arise only for $\omega_{1,1}$ and therefore $\omega_{0,4}$ is not modified. The change in $\omega_{1,2}$ comes from the following expression
\begin{equation}
	 \frac{A_m}{m} \underset{q\rightarrow t^*}{\textrm{Res}}\, q^{-2} K^{(k)}(t_1, q, \bar{q})  \omega_{0,2}(\bar{q}, t_2).
\end{equation}
The kernel has a first order pole at $q = t^*_1$ and all other terms are regular, therefore 
\begin{equation}
	-\frac{A_m}{m (t^*)^2 (t_2 - t^*)^2} \underset{q\rightarrow t^*}{\textrm{Res}} K^{(k)}(t_1, q, \bar{q}) = 
	-\frac{t^*(1-t^*)}{2m^2 n  (t_1 - t^*)^2(t_2 - t^*)^2}\frac{a_k}{(1-a_k)^2}.   \label{omega12a-uniform}
\end{equation}

We consider now contributions from $t^*_2=0$ branch point. Analyzing formulas for $\omega_{1,2}$ and $\omega_{0,4}$ and keeping in mind that $K(q,t)$ is at least linear for small $q$, we conclude that there are no extra contributions to $\omega_{0,4}$ and the potential contribution to $\omega_{1,2}$ comes only from the second term
\begin{equation}
	2 \,\underset{q\rightarrow 0}{\textrm{Res}} \,K^{(k)}(t_1, q , \bar{q}) \,\omega_{1,1}(q) \omega_{0,2}(\bar{q}, t_2) = \frac{A_m a_k}{2 m^2 t_1^2 t_2^2}.  \label{omega12b-uniform}
\end{equation}

Altogether, the extra contributions (\ref{omega12a-uniform}) and (\ref{omega12b-uniform}) to $\omega(t_1, t_2)$ are 
\begin{equation}
	\frac{A_m}{2m^2} \Big(-\frac{t^*(1-t^*)}{n  (t_1 - t^*)^2(t_2 - t^*)^2} + \sum_{\substack{k=1\\ k|f}}^{n-1}\frac{a_k}{t_1^2 t_2^2}\Big)
\end{equation}
which give rise to extra contribution to $S_3^{TR}$. The sum over the roots of unity gives zero and ultimately
\begin{equation}
%S_3^{TR} = \frac{1}{m^2} S_3^{TR,(1,mf)} + 
%	\frac{A_m}{4m^2}\Big(-\frac{t^*(1-t^*)}{n  (t - t^*)^2} + \frac{1}{t^2}\sum_{\substack{k=1\\ k|f}}^{n-1}a_k \Big),
S_3^{TR} = \frac{1}{m^2} S_3^{TR,(1,mf)} -  
	\frac{A_m}{4m^2}\frac{t^*(1-t^*)}{n  (t - t^*)^2},
\end{equation}
which is indeed consistent with (\ref{psiTR-PC-uniform}).

%***************************************************************************************************

\subsection{Quiver ${2\ 1 \brack 1\ 1}$}

In what follows we analyze several representative quivers of size $m=2$, presented in section \ref{ssec-2vertex}. The first such example is a quiver encoded in a matrix 
\be
C=\begin{bmatrix}
2 & 1 \\
1 & 1 
\end{bmatrix}  \label{C2111}
\ee
which is a quiver (\ref{family-a0}) from the family $C={a\ 0 \brack 0\ 0}$ with $a=1$ and framing $f=1$ (chosen so that there is a maximal number of ramification points). In the knots-quivers correspondence, this quiver encodes colored HOMFLY polynomials of the unknot, in framing $f = 1$ \cite{Kucharski:2017ogk}. This means that the results of this section, apart from illustrating the relation to quivers, at the same time confirm the statement that the topological recursion reconstructs colored knot polynomials.

In this case the classical resultant (\ref{resultant}) for this quiver takes form
\begin{equation}
A(x_1,x_2;y) = x_{{1}}{y}^{2}-x_{{2}}y+y-1.
\end{equation}
In this example we consider specialization $x_1 = x, x_2 = -x$, which leads to an irreducible, admissible spectral curve\footnote{The choice $x_1 = x_2 = x$ leads to a reducible spectral curve $A(x,x;y) = (y-1)(xy+1)$.}
\begin{equation}
A(x,-x,y) = x{y}^{2}+ \left( x+1 \right) y -1 = 0.  \label{sc_2111}
\end{equation}
In what follows we consider the following rational parametrization of this curve
\begin{equation}
x(t) = -{\frac {t-1}{t \left( t+1 \right) }},\qquad y(t) = t,   \label{Q211-xy-t}
\end{equation}
and the inverse function chosen to be regular at $x = 0$
\begin{equation}
t(x) = {\frac {-x-1+\sqrt {{x}^{2}+6\,x+1}}{2x}}.    \label{Q211-t-x}
\end{equation}
Furthermore, the corresponding quantum curve, which annihilates the quiver generating function with identification $x_1=-x_2=x$, takes form
\begin{equation}
\widehat{A}(x,-x,\widehat{y}) = qx\widehat{y}^{2}+\sqrt {q}x\widehat{y}+\widehat{y}-1. \label{Ahat_unkno2}
\end{equation}

We now show that the motivic generating function with the above identification of variables, as well as the above quantum curve, also arise from the topological recursion. We follow the strategy presented in previous sections. First, we suppose that the quantum curve (\ref{Ahat_unkno2}) with a simple rescaling of variables annihilates the wave function associated to the classical curve (\ref{sc_2111}) via the topological recursion
\be
\widehat{A}(q^{\alpha_1}x,-q^{\alpha_2}x,q^{\beta}\widehat{y}) \psi_{TR}(x) = 0.     \label{Q2111-AC-psi}
\ee
If we now compute $S_0^{TR}$ and $S_1^{TR}$, using (\ref{S0S1}) and the parametrization (\ref{Q211-xy-t}) and (\ref{Q211-t-x}), and substitute into the quantum curve equation (\ref{Q2111-AC-psi}), we find that it holds at the subleading order only for 
\begin{equation}
\alpha_1=\frac{3}{2},\quad \alpha_2=1,\quad \beta=-\frac12.
\end{equation}
This means that the topological recursion quantum curve should have form
\be
\widehat{A}_{TR}(\hat x,\hat y) = \widehat{A}(q^{3/2}x,-q x,q^{-1/2}\widehat{y}) = 
q^{3/2}x\widehat{y}^{2}+x\widehat{y}+q^{-1/2}\widehat{y}-1.    \label{Q2111-ATR}
\ee
We should still confirm that higher order terms of the topological recursion wave function are consistent with this result. First, substituting (\ref{Q2111-ATR}) and already determined $S_0^{TR}$ and $S_1^{TR}$ into the equation 
\be
\widehat{A}_{TR}(\hat x, \hat y) \exp\Big(  \frac{1}{\hbar}S_0^{TR} + S_1^{TR} + \sum_{k=1}^{\infty}S_k \Big)= 0,
\ee
we find the following higher order terms 
\begin{align}
\begin{split}
S_2 =& {\frac {  ( 8\,{x}^{2}X+3\,{x}^{3}+13\,xX+21\,{x}^{2}+3\,X+
		21\,x+3 ) {X}^{2}}{12  ( x+3-2\,\sqrt {2}  ) ^{3}
		 ( x+3+2\,\sqrt {2}  ) ^{3}}} , \\
S_3 =& -{\frac {  ( {x}^{2}-4\,x+11  )   ( 3\,x+X+1
		 ) x  ( x+1 ) {X}^{2}}{16  ( x+3-2\,\sqrt {2}) ^{4}  ( x+3+2\,\sqrt {2} ) ^{4}}},  \label{Q2111-S2S3S4}
%\\
%S_5 =\ & \frac{T_4}{2880\, \left( -x-3+2\,\sqrt {2} \right) ^{6} \left( x+3+2\,\sqrt {2}\right) ^{6}} \\
%S_6 =\ & \frac{T_5}{\left( x+3+2\,\sqrt {2} \right) ^{7} \left( -x-3+2\,\sqrt {2}\right) ^{7}}
\end{split}
\end{align}
where $X = \sqrt{{x}^{2}+6x+1}$. Note that the only poles of $S_k$ are at zeroes of $dx$, which is consistent with what the topological recursion should yield. 

We now show that the topological recursion correctly reproduces the above terms $S_k$. To this end, we provide first some more details concerning the curve (\ref{sc_2111}) with its parametrization (\ref{Q211-xy-t}), and the associated data. In general, the set of ramification points is given by zeroes of $dx$, and poles of $x$ of degree at least 2. Since $x$ has poles at $t = 0$ and $t = -1$, but both of them are simple, they do  not contribute to $\omega_{g,n}$. Therefore ramification points are just zeros of $dx$
\begin{equation}1+\sqrt {2},1-\sqrt {2}.
\end{equation}
The expression for the conjugate point in the neighborhoods of these two branch points is given by $\overline{t} = \frac {t+1}{t-1}$. To conduct calculations more efficiently, we write the recursion kernel as a product of a rational function in $t$ and $t_1$ and a logarithmic function
\begin{equation}
K(t,\overline{t},t_1) = {\frac {t-{t}^{3}}{ 2\left(  \left( t_{{1}}-1 \right) t-t_{{1}}-1
		\right)  \left( t_{{1}}-t \right) }} \times  \frac{1}{\ln t -\ln  {\frac {t+1}{t-1}} },
\end{equation}
and then take a product of their Laurent expansions. We then find 
\begin{equation}
\omega_{0,3}(t_1,t_2,t_3) = \frac{P_{0,3}}{Q_{0,3}},\qquad \omega_{1,1}(t_1) = \frac{P_{1,1}}{Q_{1,1}},
\end{equation}
where
\begin{equation}
\begin{aligned}
P_{0,3} =&\  (  ( 10\,{t_{{3}}}^{2}+8\,t_{{3}}+2 ) {t_{{2}}}^{2}+
( 8\,{t_{{3}}}^{2}+ 8\,t_{{3}} ) t_{{2}}+2\,{t_{{3}}}^{2}+2
) {t_{{1}}}^{2}+ \\
&\ +8\, ( t_{{3}}t_{{2}}+1 )  ( 
( t_{{3}}+1 ) t_{{2}}+t_{{3}}-1 ) t_{{1}}+ ( 2
\,{t_{{3}}}^{2}+2 ) {t_{{2}}}^{2}+ \\
&\ +( 8\,t_{{3}}-8 ) t_{{2}}+2\,{t_{{3}}}^{2}-8\,t_{{3}}+10,
\\
 Q_{0,3} = &\  ( -2+ ( t_{{1}}-1 ) \sqrt {2} )  ( t_{{1}}
-1+\sqrt {2} )  ( -1-\sqrt {2}+t_{{2}} ) ^{2} ( 
2+ ( t_{{1}}-1 ) \sqrt {2} ) \times \\
&\ \times( -1-\sqrt {2}+t_{
	{3}} ) ^{2} ( t_{{1}}-1-\sqrt {2} )  ( -1+\sqrt 
{2}+t_{{2}} ) ^{2} ( -1+\sqrt {2}+t_{{3}} ) ^{2}, 
\\
P_{1,1} =&\ 2\, ( {t_{{1}}}^{2}+1 )  ( {t_{{1}}}^{4}-16\,{t_{{1}}}
^{3}+2\,{t_{{1}}}^{2}+16\,t_{{1}}+1 )  ( {t_{{1}}}^{2}-2\,t
_{{1}}-1 ) ^{2},
\\
 Q_{1,1} =&\ -3\, ( -2+ ( t_{{1}}-1 ) \sqrt {2} ) ^{3}
( 2+ ( t_{{1}}-1 ) \sqrt {2} ) ^{3} ( t_{{
		1}}-1+\sqrt {2} ) ^{3} ( t_{{1}}-1-\sqrt {2} ) ^{3}.
\end{aligned}
\end{equation}
Computing now $S_2^{TR}$ from (\ref{psi_coeffs}), leaving the lower integration limit $t_0$ as a parameter, we get
\begin{equation}
S_2^{TR}  = {\frac { ( t_{{0}} -t)  ( tt_{{0}}+1)  }{12 ( {t_{{0}}}^{2}-2t_{{0}}-1) ^{3} ( {t}^{2}-2t-1 ) ^{3}}}  P(t,t_0),   \label{Q2111-S2TR}
\end{equation}
where
\begin{align}
\begin{split}
P(t,t_0) =&\ {t}^{4}{t_{{0}}}^{4}-10\,{t}^{4}{t_{{0}}}^{3}-10\,{t}^{3}{t_{{0}}}^{4}
-14\,{t}^{4}{t_{{0}}}^{2}+96\,{t}^{3}{t_{{0}}}^{3}-14\,{t}^{2}{t_{{0}}
}^{4}-14\,{t}^{4}t_{{0}}+\\
&\ +4\,{t}^{3}{t_{{0}}}^{2}+4\,{t}^{2}{t_{{0}}}^{
	3}-14\,t{t_{{0}}}^{4}-11\,{t}^{4}-48\,{t}^{3}t_{{0}}-4\,{t}^{2}{t_{{0}
}}^{2}-48\,t{t_{{0}}}^{3}-11\,{t_{{0}}}^{4}+\\
&\ +14\,{t}^{3}-4\,{t}^{2}t_{{0
}}-4\,t{t_{{0}}}^{2}+14\,{t_{{0}}}^{3}-14\,{t}^{2}+96\,tt_{{0}}-14\,{t
	_{{0}}}^{2}+10\,t+10\,t_{{0}}+1.
\end{split}
\end{align}
We then find that $S_2$ in (\ref{Q2111-S2S3S4}) agrees with (\ref{Q2111-S2TR}) if we fix the integration limit $t_0=1$; note that this is the zero of $x$ (and furthermore $x(t_0) = 0, y(t_0) = 1$). Similarly we computed $S_3^{TR}$, and with the same value of $t_0=1$ it reads
\be
S_3^{TR} = \frac {( 11\,{t}^{4}+26\,{t}^{3}+12\,{t}^{2}-6\,t+1) ( {t}^{2}-2\,t-1 ) ^{6}( {t}^{2}+1 ) ( t-1)^{3}
( t+1 ) t }{8( t-1-\sqrt{2}) ^{12} ( t-1+\sqrt{2} ) ^{12}},
\ee
which indeed agrees with $S_3$ in (\ref{Q2111-S2S3S4}).

To sum up, we conclude that the topological recursion determines (perturbatively in $\hbar$) the motivic generating series for the quiver encoded by the matrix (\ref{C2111}), as well as the corresponding quantum curve (\ref{Ahat_unkno2}).

%***************************************************************************************************

\subsection{Quiver ${3\ 1 \brack 1\ 1}$} \label{ssec-3111}    

We now discuss another quiver from the family ${a\ 0 \brack 0\ 0}$ in (\ref{family-a0}), this time with $a=2$ and $f=1$
\be
C=\begin{bmatrix}
3 & 1 \\
1 & 1 
\end{bmatrix}  \label{C3111}
\ee
We recall that this quiver encodes extremal colored polynomials of the left-handed trefoil knot in framing $f = 1$, and also (in another framing) counting of Duchon paths \cite{Panfil:2018sis}. This means that the results of this section, apart from illustrating the relation to quivers, at the same time confirm the statement that the topological recursion reconstructs colored knot polynomials \cite{DijkgraafFuji-2,Borot:2012cw,abmodel}, as well as counts of lattice paths.

We have already presented various results associated to quivers of the form ${a\ 0 \brack 0\ 0}$ in section \ref{ssec-2vertex}. For $a=2$ and framing $f=1$, quantum A-polynomial takes form
\begin{equation}
    \widehat{A}(x_1,x_2,\widehat{y}) = q^{\frac{3}{2}}x_1\widehat{y}^3 + (q^3x_2^2-q^{\frac{3}{2}}x_2)\widehat{y}^2+((q+1)\sqrt{q}x_2 -1)\widehat{y} + 1,
\end{equation}
the classical A-polynomial, with identification $x_1=x_2=x$, reads
\begin{equation}
    A(x,y) = xy^3+(x^2-x)y^2+(2x-1)y+1,
\end{equation}
and its parametrization takes form (\ref{param-a000})
\begin{equation}
x(t) = \frac{t+1}{t(t^2-t-1)},\quad y(t) = \frac{t^2(t-1)}{t^2-t-1}.
\end{equation}
Comparing the subleading term $S_1$ from the saddle point expansion of the quiver generating series, with the topological recursion result, similarly as in earlier examples we find the quantum curve that annihilates topological recursion partition function
\begin{equation}
    \widehat{A}_{TR}(\hat x, \hat y) = \widehat{A}(q^2x,qx,q^{-\frac{1}{2}}\widehat{y}) = q^2x_1\widehat{y}^3 + (q^4x_2^2-q^{\frac{3}{2}}x_2)\widehat{y}^2+((q+1)qx_2 -q^{-\frac{1}{2}})\widehat{y} + 1.
\end{equation}
In this case we cannot write analytically higher subleading terms. Nonetheless, the above result confirms that the topological recursion formalism reconstructs the quiver generating function at the subleading order, and from consistency of the topological recursion we expect that it holds to all orders.

%***************************************************************************************************

\subsection{Quiver ${2\ \ 0 \brack 0\, -1}$}

The next example we consider is a reciprocal quiver
\begin{equation}\label{b-family}
\begin{bmatrix}
2 & 0 \\
0 & -1 
\end{bmatrix}
\end{equation}
which is a representative of the family (\ref{family-a1-a}) with $a=2$, presented in section \ref{ssec-2vertex}. The operator that annihilates the corresponding quantum generating function takes form
\be
\widehat{A}(x_1,x_2,\widehat{y}) = {q}^{3}x_1^{2}{\widehat{y}}^{4}+x_{{1}}{\widehat{y}}^{3}- \big( {q}^{3/2} ( q+1 ) x_{{1}}x_{{2}}+x_{{1}}+\sqrt {q}x_{{2}} \big) {\widehat{y}}^{2}+
\sqrt {q}x_{{2}}\widehat{y}+qx_2^2.
\ee
In what follows we consider the specialization $x_1 = x$ and $x_2 = -x$, which leads to the classical spectral curve of the form
\begin{equation}
A(x,-x,y) = x ( x{y}^{4}+{y}^{3}+2\,x{y}^{2}-y+x ) = 0.
\end{equation}
Apparently, the prefactor $x$ does not affect the topological recursion calculation, and the non-trivial factor in the bracket gives an irreducible curve of genus zero. We consider the following parametrization of this curve
\begin{equation}
x(t) = -{\frac { ( t-1 ) ( t+1 ) t}{ ( {t}^{2}+1 ) ^{2}}},\qquad y(t) = t.   \label{C200-1-param}
\end{equation}
Proceeding as in previous sections, having computed $S_0^{TR}$ and $S_1^{TR}$ as in (\ref{S0S1}), we find that the topological recursion quantum curve is expected to take form
\begin{equation}
\widehat{A}_{TR} = {q}^{3}{x}^{2}{\widehat{y}}^{4}+q^{-1}x\widehat{y}^{3}+{q}^{3/2} \left( q+1
\right) {x}^{2}{\widehat{y}}^{2}-x\widehat{y}+q{x}^{2}.
\end{equation}
Substituting this result to the equation
\be
\widehat{A}_{TR}(\hat x, \hat y) \exp\Big(  \frac{1}{\hbar}S_0^{TR} + S_1^{TR} + \sum_{k=1}^{\infty}S_k \Big)= 0,   \label{C200-1-ATR}
\ee
we find
\begin{align}
\begin{split}
S_2 = &-{\frac { ( x-1 ) ( 144{x}^{3}+144{x}^{2}+11x+
		11) }{60( 16{x}^{2}-1 ) ^{2}}}-{\frac {x( 368{x}^{2}+77) X}{48( 16\,{x}^{2}-1) ^{2}}}+ \\
&\ + {\frac { ( 448{x}^{4}-540{x}^{2}+7) {X}^{2}}{48( 16{x}^{2}-1) ^{2}}}-{\frac { ( 432{x}^{2}-7 ) x{X}^{3}}{48 ( 16{x}^{2}-1 ) ^{2}}},    \label{200-1s2wkb}
\end{split}
\end{align}
where $X$ is a solution of the equation $x{X}^{4}+{X}^{3}+2\,x{X}^{2}-X+x = 0$.

We now confirm that the same $S_2$ follows from the topological recursion calculation. From the parametrization (\ref{C200-1-param}) we find the ramification points
\begin{equation}
\sqrt {2}-1,-1-\sqrt {2},1-\sqrt {2},1+\sqrt {2},  \label{C200-1-Ram}
\end{equation}
and the following two kernels associated to appropriate ramification points
\begin{align}
\begin{split}
K_{-1\pm\sqrt{2}} =&-{\frac {t( t-1)( t+1)( {t}^{2}+1) }{2( {t}^{2}-2t-1)( t_1-t) ( p_{{1}}t+t_1+t-1) }} \times \frac{1}{\ln t -\ln( -t+1) +\ln( t+1)}, \\
K_{1\pm\sqrt{2}} =&-{\frac {t ( t-1)( t+1)( {t}^{2}+1) }{2 ( {t}^{2}+2t-1)( t_1-t)( p_{{1}}t-t_1-t-1)} \times \frac{1}{\ln t -\ln( t+1) +\ln( t-1) } }.
\end{split}
\end{align}
Interestingly, in this case $\omega_{0,3}$ and $\omega_{1,1}$ are equal to those for the quiver $((2,0),(0,2))$ in section \ref{ssec-2002} (if we do not consider points at infinity). In the present case, poles of $x$ (residues at infinity) do not contribute, since they cannot be canceled by zeros of $y$.
In the present case we find that the integration limit is $t_0=1$, so that $x(t_0) = 0$ and $y(t_0) = 1$.
% however this point it is different than in section \ref{ssec-2002}. 
 It then follows that
\begin{align}
\begin{split}
S_2^{TR} & = {\frac {  (t-1)^{2}}{48( t-1+\sqrt {2}) ^{3} ( -t+\sqrt {2}-1 ) ^{3} ( -t+1+\sqrt {2}
 ) ^{3} ( t+1+\sqrt {2} ) ^{3}}}  \label{200-1s2tr} \times \\
 & \times ( 17{t}^{10}+132{t}^{9}+181{t}^{8}+144{t}^
{7}-118{t}^{6}-232{t}^{5}-118{t}^{4}+144{t}^{3}+181{t}^{2}+
132t+17 ) 
\end{split}
\end{align}
which agrees with the result (\ref{200-1s2wkb}). We conclude that the topological recursion reproduces the quiver generating function in this case too.

%***************************************************************************************************

\subsection{Quiver ${2\ 2 \brack 2\ 1}$}

Another quiver that we consider is a representative of the family introduced in section \ref{ssec-family} that takes form (\ref{family-ccc0}), with $c=1$ and framing $f=1$
\be
\begin{bmatrix}
2 & 2 \\
2 & 1 
\end{bmatrix}
\ee
The quantum Nahm equations take form
\begin{align}
\begin{split}
     \hat{z}_1 P(x_1,x_2) & =  (1- q x_1 \hat{y}) P(x_1,x_2), \\
     \hat{z}_2 P(x_1,x_2) & =  (1 + x_2 \hat{y} (1 - x_1 \hat{y})) P(x_1,x_2),
\end{split}
\end{align}
so that the quantum operator that annihilates the quantum generating function takes form
\be
\hat{A}(x_1,x_2,\hat{y}) = q^{15/2} x_1^2 x_2 \hat{y}^5 - 2 q^{5/2} x_1 x_2 \hat{y}^3 - q x_1 \hat{y}^2 + (q^{1/2} x_2 - 1) \hat{y} + 1, 
\ee
and its classical limit 
\be
A(x_1,x_2,y) = x_1^2 x_2 y^5 - 2 x_1 x_2 y^3 - x_1 y^2 + (x_2 - 1) y + 1.
\ee
Let us consider the specialization $x_1=x_2=x$, for which the A-polynomial reads\footnote{
Notice that specialization $x_1=-x_2=x$ gives a reducible curve $$A(x,-x,y) = ( x{y}^{2}-1)\big(  ( xy ( x{y}^{2}-1 ) +1 ) +y\big) .$$}
\begin{equation}
A(x,x,y) %= (xy^2-1)(xy(xy^2-1)-1)-y  
= x^3 y^5 - 2 x^2 y^3 - x y^2 + (x - 1) y + 1 = 0,
\end{equation}
and whose parametrization takes form
\begin{equation}
x(t) = {\frac {{t}^{4}}{ ( t+1 )  ( {t}^{2}-t-1 ) ^{2}}}, \qquad
y(t) = {\frac { ( t+1 )  ( {t}^{2}-t-1) }{{t}^{3}}}.   \label{C2221-param}
\end{equation}
The quantum A-polynomial that annihilates the corresponding generating function takes form
\begin{equation}
\widehat{A}(x,x,y) = {q}^{15/2}{x}^{3}\widehat{y}^{5}- ( q+1 ) {q}^{5/2}{x}^{2}\widehat{y}^{3}-q
x\widehat{y}^{2}+\sqrt {q}x\widehat{y}-\widehat{y} +1.
\end{equation}

Because the A-polynomial is given by the quintic equation, we cannot write down explicitly exact expressions in topological recursion calculation. Nonetheless, we can make the basic check that the subleading term is correct. Computing $S_1^{TR}$ as in (\ref{S0S1}), using the parametrization (\ref{C2221-param}), and substituting the result to
\be
\widehat{A}(q^{\alpha_1}x,q^{\alpha_2}x,q^{\beta}\widehat{y}) \psi_{TR}(x) = 0,     \label{Q2221-AC-psi}
\ee
we find that there is a unique solution 
\be
\alpha_1=-\frac12,\qquad \alpha_2=0,\qquad \beta=\frac12,
\ee
which confirms the agreement at the subleading order. The above solution implies that the topological quantum curve takes form
\begin{equation}
\widehat{A}_{TR} = {q}^{9}{x}^{3}\widehat{y}^{5}- \left( q+1 \right) {q}^{7/2}{x}^{2}\widehat{y}^{3}-{q}^
{3/2}x\widehat{y}^{2}+qx\widehat{y}-q^{1/2}\widehat{y}+1,
\end{equation}
and we expect that higher order $S_k^{TR}$ terms should agree with this form.

%***************************************************************************************************

\subsection{Quiver ${3\ 2 \brack 2\ 1}$}   \label{ssec-3221}

Yet another example of a quiver from the family presented in section \ref{ssec-family} is that of the form (\ref{family-c+1cc0}), with $c=1$ and farming $f=1$
\begin{equation}
\begin{bmatrix}
3 & 2 \\
2 & 1
\end{bmatrix}
\end{equation}
We remind that the data associated to this quiver captures extremal invariants of the right-handed trefoil knot and (after adjusting framing) counting of certain Duchon paths \cite{Panfil:2018sis}. Therefore the results of this section support the conjecture that relates knot invariants and the topological recursion  \cite{DijkgraafFuji-2,Borot:2012cw,abmodel}, and reveal novel links between path counting problems and the topological recursion.

In this case the quantum A-polynomial takes form
\be
\widehat{A}(x_1,x_2;\widehat{y}) = {q}^{4}{x_1}^{2}\widehat{y}^{5}- \left( q+1 \right) \sqrt {q}x_{{1}}\widehat{y}^{3}+\sqrt {q}x_{{1}}\widehat{y}^{2}+ \left( -\sqrt {q}x_{{2}}+1 \right) \widehat{y}-1.
\ee
With identification $x_1=x_2=x$ the classical A-polynomials reads
\be
A(x,x;y) = y(xy^2-1)^2+xy(y-1)-1,
\ee
and has the following parametrization 
\be
x(t)=-{\frac { \left( {t}^{2}+t-1 \right) ^{2}}{{t}^{4} \left( t-1\right) }},\qquad y(t)={\frac {t \left( t-1 \right) }{{t}^{2}+t-1}}.
\ee
In this case it is not possible to write down exact results for higher order terms. However we can match the quiver generating function and the topological recursion wave-function at the subleading order, which is already a non-trivial check. We find that this matching arises for $x_1 = q^a x,\ x_2 = q^b x, y = q^c y$ with $(a,b,c) = (-1,0,1/2)$, and then the topological recursion quantum curve reads
\begin{equation}
\widehat{A}_{TR} = {q}^{9/2}{x}^{2}\widehat{y}^{5}- \left( q+1 \right) qx\widehat{y}^{3}+\sqrt {q}x\widehat{y}^{2
}+ \left( -qx+\sqrt {q} \right) \widehat{y}-1
\end{equation}
This confirms our postulate at the subleading order.
%From the WKB for $\widehat{A}_{\mathrm{TR}}$:
%\begin{equation}
%S_2(x)^{\mathrm{WKB}} = const. + -\frac{x}{12}+{\frac {13\,{x}^{2}}{6}}+{\frac {475\,{x}^{3}}{8}}+{\frac {2990\,{x}^{4}}{3}}+{\frac {112915\,{x}^{5}}{8}} + \dots
%\end{equation}

%***************************************************************************************************

\subsection{Quiver ${2\ 0 \brack 0\ 2}$}   \label{ssec-2002}

At the end we consider a diagonal quiver, introduced earlier in section \ref{ssec-2vertex}. Here we come across one unfavorable result -- namely, in the quantum curve determined by the topological recursion one term appears to arise in different ordering than in the operator that annihilates the quiver generating function. While we believe that our conjecture is in general true, it may happen that for diagonal quivers some subtlety arises, that deserves further analysis. Nonetheless, this example of diagonal quivers is quite instructive in itself, so we briefly summarize it. The diagonal quiver that we consider has the form (\ref{diagonal-22})
\be
\begin{bmatrix} 
2 & 0 \\
0 & 2 \\ 
\end{bmatrix}    \label{C2002}
\ee
Various quantities associated to this quiver, in particular classical and quantum A-polynomials, have been already presented in section \ref{ssec-2vertex}.
%\be
%\hat{A}(x_1,x_2,\hat{y}) = q^9 x_1^2 x_2^2 \hat{y}^4 - q^3 x_1 x_2 \hat{y}^3 - q(x_1 + x_2 + q(q + 1) x_1 x_2) \hat{y}^2 - \hat{y} + 1,
%\ee
%and its classical limit, or equivalently the classical resultant, takes form
%\be
%A(x_1,x_2,y) = y^4 x_1^2 x_2^2 - y^3 x_1 x_2 - (x_1 + x_ 2 + 2 x_1 x_2) y^2 - y + 1 .
%\ee
Here we are interested in the specialization $x_1=-x,x_2=x$ which yields the following quantum curve
\begin{equation}
\widehat{A}(x,\widehat{y}) = {q}^{9}{x}^{4}\widehat{y}^{4}+{q}^{3}{x}^{2}\widehat{y}^{3}+ \left( q+1 \right) {q}^{2}{x}^{2}\widehat{y}^{2}-\widehat{y}+1,    \label{AhaQ_2002}
\end{equation}
and an admissible classical curve of genus zero\footnote{The specialization $x_1=x_2=x$ leads to a factorizable curve $A(x,x,y) = \left( {x}^{2}{y}^{2}-2\,xy-y+1 \right)  \left( xy+1 \right) ^{2}$.}
\begin{equation}
A(x,y) = {x}^{4}{y}^{4}+{x}^{2}{y}^{3}+2\,{x}^{2}{y}^{2}-y+1 = 0.   \label{C2002-class}
\end{equation}
This curve has 4 branches, therefore we can introduce 4 parametrizations related by transformations that exchange these branches. For $x = \frac{(t-1)(t+1)t}{(t^2+1)^2}$, the $y$ coordinate for those different branches is parametrized by 
\be
y_1 ={\frac { \left( {t}^{2}+1 \right) ^{2}}{{t}^{2} \left( t-1 \right)  \left( t+1 \right) }}, \qquad
y_2 =-{\frac { \left( {t}^{2}+1 \right) ^{2}}{ \left( t-1 \right)  \left( t+1 \right) }}, \qquad
y_3 ={\frac { \left( {t}^{2}+1 \right) ^{2}}{ \left( t+1 \right) ^{2}t}},\qquad
y_4 =-{\frac { \left( {t}^{2}+1 \right) ^{2}}{ \left( t-1 \right) ^{2}t}}.
\ee
Computing now $S_1^{TR}$ as in (\ref{S0S1}), and taking into account some features of $S_2^{TR}$ we find that in those cases the corresponding quantum curves would take form
\begin{align}
\begin{split}
\widehat{A}^{(1)}_{TR}(\hat x,\hat y) &={q}^{10}{x}^{4}{\hat y}^{4}+{q}^{7/2}{x}^{2}{\hat y}^{3}+ \left( q+1 \right) {q}
^{5/2}{x}^{2}{\hat y}^{2}+1-\sqrt {q}\hat  y, \\
\widehat{A}^{(2)}_{TR}(\hat x,\hat y) &={q}^{10}{x}^{4}{\hat y}^{4}+{q}^{5/2}{x}^{2}{\hat y}^{3}+ \left( q+1 \right) {q}
^{5/2}{x}^{2}{\hat y}^{2}+1-{\frac {\hat y}{\sqrt {q}}}, \\
\widehat{A}^{(3)}_{TR}(\hat x,\hat y) &={q}^{10}{x}^{4}{\hat y}^{4}+{q}^{7/2}{x}^{2}{\hat y}^{3}+ \left( q+1 \right) {q}
^{5/2}{x}^{2}{\hat y}^{2}-{\frac {\hat y}{\sqrt {q}}}+ \underline{q^{1/2}}\left( q-1 \right) x{\hat y}^{2}+1, \\
\widehat{A}^{(4)}_{TR}(\hat x,\hat y) &={q}^{10}{x}^{4}{\hat y}^{4}+{q}^{7/2}{x}^{2}{\hat y}^{3}+ \left( q+1 \right) {q}
^{5/2}{x}^{2}{\hat y}^{2}-{\frac {\hat y}{\sqrt {q}}}+ \underline{q^{3/2}}\left( {q}^{-1}-1 \right) x{\hat y}^{2}+1.
\end{split}  \label{AhatC-2002}
\end{align}
The underlined monomial coefficients are necessary and defined uniquely from $S_2$ term (if they are not included, then some unwanted logarithmic additional terms appear in higher $S_k$'s). The parametrizations $(x,y_1)$ and $(x,y_2)$ are related by $t\mapsto \frac{1}{t}$, while $(x,y_3)$ and $(x,y_4)$ are related by $t\mapsto -\frac{1}{t}$. Note that for $(x,y_3)$ and $(x,y_4)$, the underlined term $(q^{\pm 1}-1)xy^2$ vanishes when $q\rightarrow 1$, which means that it is non-trivially generated by quantum effects.

Now we would like to match either of the topological recursion quantum curves in (\ref{AhatC-2002}) with the operator (\ref{AhaQ_2002}). We find that
\begin{equation}
\widehat{A}^{(1)}_{TR}(q^{1/4}x,q^{-1/2}\hat y) = \widehat{A}^{(2)}_{TR}(q^{-3/4}x,q^{1/2}\hat y) = {q}^{9}{x}^{4}{\hat y}^{4}+{q}^{5/2}{x}^{2}{\hat y}^{3}+ \left( q+1 \right) {q}^{2}{x}^{2}{\hat y}^{2}-{\hat y} +1. \label{2002_atr1_atr_2_rescaled}
\end{equation}
Therefore the difference between the required form (\ref{AhatC-2002}) appears only in the second monomial: instead its prefactor $q^{5/2}$ we would expect the prefactor $q^3$. Note that the missing $q^{1/2}$ can be taken care of by changing the ordering of operators, and writing this monomial as $q^{5/2}x^2\hat y^3 = q^3 x({\hat y}^{-1/2}x{\hat y}^{1/2})\hat y^3$. We see that the topological recursion seems to be missing the correct result by a seemingly minor factor, whose origin we have unfortunately not been able to identify.

%We therefore proceed to the WKB solutions to compare the $S$ coefficients from (\ref{2002_atr1_atr_2_rescaled}) and quiver curve, to see how the change of a single monomial affects the WKB solution.
%\begin{equation}
%\begin{aligned}
%S_1^{Q} =&\ \frac{1}{4}(\,\ln  \left( {X}^{3}x+ \left( -{x}^{2}
%-1 \right) {X}^{2}+3\,Xx-2\,{x}^{2}+1 \right) -\,\ln  \left( 4\,x-1
% \right) -\,\ln  \left( 4\,x+1 \right)) + \\
% &\ \frac{1}{2}\,\ln  \left( x \right) \\
%S_1^{TR} =&\ \frac{1}{4}(\ln  \left( {X}^{3}x+ \left( -{x}^{2}-1 \right) {X}^{2}+3\,Xx-3\,{x}^{
%2}+1 \right) -\ln  \left( 4\,x-1 \right) -\ln  \left( 4\,x+1 \right)) \\
%\end{aligned}
%\end{equation}
%Here, under the first logarithm coefficients in front of $x^2$ differ, which will affect all other $S_k$ as well. But $S_0$ of course agree since the classical limit is the same.
%----

%***************************************************************************************************
%***************************************************************************************************
%***************************************************************************************************

%\newpage

\acknowledgments{We thank Bertrand Eynard, Sergei Gukov, Piotr Kucharski and Marko Sto$\check{\text{s}}$i$\acute{\text{c}}$ for discussions, correspondence, and comments on the manuscript. This work has been supported by the ERC Starting Grant no. 335739 \emph{``Quantum fields and knot homologies''} funded by the European Research Council under the European Union's Seventh Framework Programme, and the TEAM programme of the Foundation for Polish Science co-financed by the European Union under the European Regional Development Fund (POIR.04.04.00-00-5C55/17-00). MP acknowledges the support from the National Science Centre, Poland, in the initial phase of the project under the FUGA grant 2015/16/S/ST2/00448 and in the final phase of the project under the SONATA grant 2018/31/D/ST3/03588.}

%************************************************************************************
%************************************************************************************
%************************************************************************************

\newpage

\bibliographystyle{JHEP}
\bibliography{abmodel}

\end{document}